\documentclass[12pt,english]{article}
\usepackage{avant}
\usepackage[T1]{fontenc}
\usepackage[latin9]{inputenc}
\usepackage{geometry}
\usepackage{booktabs}
\usepackage{babel}
\usepackage{array}
\usepackage{float}
\usepackage{multirow}
\usepackage{amsmath}
\usepackage{amssymb}
\usepackage{graphicx}
\usepackage[authoryear]{natbib}
\usepackage[unicode=false,
 bookmarks=false, breaklinks=false,pdfborder={0 0 1},colorlinks=false]{hyperref}

\makeatletter

\providecommand{\tabularnewline}{\\}
\floatstyle{ruled}
\newfloat{algorithm}{tbp}{loa}
\providecommand{\algorithmname}{Algorithm}
\floatname{algorithm}{\protect\algorithmname}

\@ifundefined{date}{}{\date{}}

\DeclareMathOperator*{\argmax}{argmax}

\DeclareMathOperator{\IF}{IF}

\usepackage{babel}
\usepackage{array}
\newcolumntype{L}[1]{>{\raggedright\let\newline\\\arraybackslash\hspace{0pt}}m{#1}}
\newcolumntype{C}[1]{>{\centering\let\newline\\\arraybackslash\hspace{0pt}}m{#1}}
\newcolumntype{R}[1]{>{\raggedleft\let\newline\\\arraybackslash\hspace{0pt}}m{#1}}
\usepackage{multirow}

\providecommand{\tabularnewline}{\\}
\floatstyle{ruled}
\newfloat{algorithm}{tbp}{loa}
\providecommand{\algorithmname}{Algorithm}
\floatname{algorithm}{\protect\algorithmname}

\usepackage{amsthm}\usepackage{bm}\usepackage{rotating}\usepackage{array}\usepackage{booktabs}\usepackage{color}\usepackage{algorithmic}
\usepackage{algorithm}\usepackage{multirow}\usepackage{enumerate}\usepackage{caption}\usepackage{subcaption}

\floatstyle{ruled}
\newfloat{algorithm}{tbp}{loa}
\providecommand{\algorithmname}{Algorithm}
\floatname{algorithm}{\protect\algorithmname}

\topmargin by -\headheight \advance
\topmargin by -\headsep

\evensidemargin
\oddsidemargin
\providecommand{\tabularnewline}{\\}

\@ifundefined{definecolor}{color}{}
\RequirePackage[displaymath]{lineno}

\@ifundefined{definecolor}{color}{}
\usepackage{amsfonts}\usepackage{latexsym}

\usepackage{mathrsfs}\usepackage{amsthm}\usepackage{latexsym}\usepackage{booktabs}\@ifundefined{definecolor}{color}{}

\usepackage{times}\@ifundefined{definecolor}{color}{}

\newtheorem{lem}{{\bf Lemma}}\newtheorem{thm}{{\bf Theorem}}

\usepackage{babel}

\makeatother

\begin{document}
{\setlength{\baselineskip}{1.5\baselineskip} 
\global\long\def\mba{\mathbf{a}}%
\global\long\def\mbA{\mathbf{A}}%
\global\long\def\mbb{\mathbf{b}}%
\global\long\def\mbB{\mathbf{B}}%
\global\long\def\mbc{\mathbf{c}}%
\global\long\def\mbC{\mathbf{C}}%
\global\long\def\mbd{\mathbf{d}}%
\global\long\def\mbD{\mathbf{D}}%
\global\long\def\mbe{\mathbf{e}}%
\global\long\def\mbE{\mathbf{E}}%
\global\long\def\mbf{\mathbf{f}}%
\global\long\def\mbF{\mathbf{F}}%
\global\long\def\mbg{\mathbf{g}}%
\global\long\def\mbG{\mathbf{G}}%
\global\long\def\mbh{\mathbf{h}}%
\global\long\def\mbH{\mathbf{H}}%
\global\long\def\mbi{\mathbf{i}}%
\global\long\def\mbI{\mathbf{I}}%
\global\long\def\mbj{\mathbf{j}}%
\global\long\def\mbJ{\mathbf{J}}%
\global\long\def\mbk{\mathbf{k}}%
\global\long\def\mbK{\mathbf{K}}%
\global\long\def\mbl{\mathbf{l}}%
\global\long\def\mbL{\mathbf{L}}%
\global\long\def\mbm{\mathbf{m}}%
\global\long\def\mbM{\mathbf{M}}%
\global\long\def\mbn{\mathbf{n}}%
\global\long\def\mbN{\mathbf{N}}%
\global\long\def\mbo{\mathbf{o}}%
\global\long\def\mbO{\mathbf{O}}%
\global\long\def\mbp{\mathbf{p}}%
\global\long\def\mbP{\mathbf{P}}%
\global\long\def\mbq{\mathbf{q}}%
\global\long\def\mbQ{\mathbf{Q}}%
\global\long\def\mbr{\mathbf{r}}%
\global\long\def\mbR{\mathbf{R}}%
\global\long\def\mbs{\mathbf{s}}%
\global\long\def\mbS{\mathbf{S}}%
\global\long\def\mbt{\mathbf{t}}%
\global\long\def\mbT{\mathbf{T}}%
\global\long\def\mbu{\mathbf{u}}%
\global\long\def\mbU{\mathbf{U}}%
\global\long\def\mbv{\mathbf{v}}%
\global\long\def\mbV{\mathbf{V}}%
\global\long\def\mbw{\mathbf{w}}%
\global\long\def\mbW{\mathbf{W}}%
\global\long\def\mbx{\mathbf{x}}%
\global\long\def\mbX{\mathbf{X}}%
\global\long\def\mby{\mathbf{y}}%
\global\long\def\mbY{\mathbf{Y}}%
\global\long\def\mbz{\mathbf{z}}%
\global\long\def\mbZ{\mathbf{Z}}%
\global\long\def\hatmba{\widehat{\mathbf{a}}}%
\global\long\def\hatmbA{\widehat{\mathbf{A}}}%
\global\long\def\hatmbb{\widehat{\mathbf{b}}}%
\global\long\def\hatmbB{\widehat{\mathbf{B}}}%
\global\long\def\hatmbc{\widehat{\mathbf{c}}}%
\global\long\def\hatmbC{\widehat{\mathbf{C}}}%
\global\long\def\hatmbd{\widehat{\mathbf{d}}}%
\global\long\def\hatmbD{\widehat{\mathbf{D}}}%
\global\long\def\hatmbe{\widehat{\mathbf{e}}}%
\global\long\def\hatmbE{\widehat{\mathbf{E}}}%
\global\long\def\hatmbf{\widehat{\mathbf{f}}}%
\global\long\def\hatmbF{\widehat{\mathbf{F}}}%
\global\long\def\hatmbg{\widehat{\mathbf{g}}}%
\global\long\def\hatmbG{\widehat{\mathbf{G}}}%
\global\long\def\hatmbh{\widehat{\mathbf{h}}}%
\global\long\def\hatmbH{\widehat{\mathbf{H}}}%
\global\long\def\hatmbi{\widehat{\mathbf{i}}}%
\global\long\def\hatmbI{\widehat{\mathbf{I}}}%
\global\long\def\hatmbj{\widehat{\mathbf{j}}}%
\global\long\def\hatmbJ{\widehat{\mathbf{J}}}%
\global\long\def\hatmbk{\widehat{\mathbf{k}}}%
\global\long\def\hatmbK{\widehat{\mathbf{K}}}%
\global\long\def\hatmbl{\widehat{\mathbf{l}}}%
\global\long\def\hatmbL{\widehat{\mathbf{L}}}%
\global\long\def\hatmbm{\widehat{\mathbf{m}}}%
\global\long\def\hatmbM{\widehat{\mathbf{M}}}%
\global\long\def\hatmbn{\widehat{\mathbf{n}}}%
\global\long\def\hatmbN{\widehat{\mathbf{N}}}%
\global\long\def\hatmbo{\widehat{\mathbf{o}}}%
\global\long\def\hatmbO{\widehat{\mathbf{O}}}%
\global\long\def\hatmbp{\widehat{\mathbf{p}}}%
\global\long\def\hatmbP{\widehat{\mathbf{P}}}%
\global\long\def\hatmbq{\widehat{\mathbf{q}}}%
\global\long\def\hatmbQ{\widehat{\mathbf{Q}}}%
\global\long\def\hatmbr{\widehat{\mathbf{r}}}%
\global\long\def\hatmbR{\widehat{\mathbf{R}}}%
\global\long\def\hatmbs{\widehat{\mathbf{s}}}%
\global\long\def\hatmbS{\widehat{\mathbf{S}}}%
\global\long\def\hatmbt{\widehat{\mathbf{t}}}%
\global\long\def\hatmbT{\widehat{\mathbf{T}}}%
\global\long\def\hatmbu{\widehat{\mathbf{u}}}%
\global\long\def\hatmbU{\widehat{\mathbf{U}}}%
\global\long\def\hatmbv{\widehat{\mathbf{v}}}%
\global\long\def\hatmbV{\widehat{\mathbf{V}}}%
\global\long\def\hatmbw{\widehat{\mathbf{w}}}%
\global\long\def\hatmbW{\widehat{\mathbf{W}}}%
\global\long\def\hatmbx{\widehat{\mathbf{x}}}%
\global\long\def\hatmbX{\widehat{\mathbf{X}}}%
\global\long\def\hatmby{\widehat{\mathbf{y}}}%
\global\long\def\hatmbY{\widehat{\mathbf{Y}}}%
\global\long\def\hatmbz{\widehat{\mathbf{z}}}%
\global\long\def\hatmbZ{\widehat{\mathbf{Z}}}%
\global\long\def\bolalpha{\boldsymbol{\alpha}}%
\global\long\def\bolbeta{\boldsymbol{\beta}}%
\global\long\def\bolgamma{\boldsymbol{\gamma}}%
\global\long\def\boldelta{\boldsymbol{\delta}}%
\global\long\def\bolepsilon{\boldsymbol{\epsilon}}%
\global\long\def\bolvarepsilon{\boldsymbol{\varepsilon}}%
\global\long\def\bolzeta{\boldsymbol{\zeta}}%
\global\long\def\boleta{\boldsymbol{\eta}}%
\global\long\def\boltheta{\boldsymbol{\theta}}%
\global\long\def\bolkappa{\boldsymbol{\kappa}}%
\global\long\def\bollambda{\boldsymbol{\lambda}}%
\global\long\def\bolmu{\boldsymbol{\mu}}%
\global\long\def\bolnu{\boldsymbol{\nu}}%
\global\long\def\bolxi{\boldsymbol{\xi}}%
\global\long\def\bolpi{\boldsymbol{\pi}}%
\global\long\def\bolrho{\boldsymbol{\rho}}%
\global\long\def\bolsigma{\boldsymbol{\sigma}}%
\global\long\def\boltau{\boldsymbol{\tau}}%
\global\long\def\bolphi{\boldsymbol{\phi}}%
\global\long\def\bolchi{\boldsymbol{\chi}}%
\global\long\def\bolpsi{\boldsymbol{\psi}}%
\global\long\def\bolomega{\boldsymbol{\omega}}%
\global\long\def\bolGamma{\boldsymbol{\Gamma}}%
\global\long\def\bolDelta{\boldsymbol{\Delta}}%
\global\long\def\bolTheta{\boldsymbol{\Theta}}%
\global\long\def\bolLambda{\boldsymbol{\Lambda}}%
\global\long\def\bolPi{\boldsymbol{\Pi}}%
\global\long\def\bolSigma{\boldsymbol{\Sigma}}%
\global\long\def\bolPhi{\boldsymbol{\Phi}}%
\global\long\def\bolPsi{\boldsymbol{\Psi}}%
\global\long\def\bolOmega{\boldsymbol{\Omega}}%
\global\long\def\hatbolalpha{\widehat{\boldsymbol{\alpha}}}%
\global\long\def\hatbolbeta{\widehat{\boldsymbol{\beta}}}%
\global\long\def\hatbolgamma{\widehat{\boldsymbol{\gamma}}}%
\global\long\def\hatboldelta{\widehat{\boldsymbol{\delta}}}%
\global\long\def\hatbolepsilon{\widehat{\boldsymbol{\epsilon}}}%
\global\long\def\hatbolzeta{\widehat{\boldsymbol{\zeta}}}%
\global\long\def\hatboleta{\widehat{\boldsymbol{\eta}}}%
\global\long\def\hatboltheta{\widehat{\boldsymbol{\theta}}}%
\global\long\def\hatbolkappa{\widehat{\boldsymbol{\kappa}}}%
\global\long\def\hatbollambda{\widehat{\boldsymbol{\lambda}}}%
\global\long\def\hatbolmu{\widehat{\boldsymbol{\mu}}}%
\global\long\def\hatbolnu{\widehat{\boldsymbol{\nu}}}%
\global\long\def\hatbolxi{\widehat{\boldsymbol{\xi}}}%
\global\long\def\hatbolpi{\widehat{\boldsymbol{\pi}}}%
\global\long\def\hatbolrho{\widehat{\boldsymbol{\rho}}}%
\global\long\def\hatbolsigma{\widehat{\boldsymbol{\sigma}}}%
\global\long\def\hatboltau{\widehat{\boldsymbol{\tau}}}%
\global\long\def\hatbolphi{\widehat{\boldsymbol{\phi}}}%
\global\long\def\hatbolchi{\widehat{\boldsymbol{\chi}}}%
\global\long\def\hatbolpsi{\widehat{\boldsymbol{\psi}}}%
\global\long\def\hatbolomega{\widehat{\boldsymbol{\omega}}}%
\global\long\def\hatbolGamma{\widehat{\boldsymbol{\Gamma}}}%
\global\long\def\hatbolDelta{\widehat{\boldsymbol{\Delta}}}%
\global\long\def\hatbolTheta{\widehat{\boldsymbol{\Theta}}}%
\global\long\def\hatbolLambda{\widehat{\boldsymbol{\Lambda}}}%
\global\long\def\hatbolPi{\widehat{\boldsymbol{\Pi}}}%
\global\long\def\hatbolSigma{\widehat{\boldsymbol{\Sigma}}}%
\global\long\def\hatbolPhi{\widehat{\boldsymbol{\Phi}}}%
\global\long\def\hatbolPsi{\widehat{\boldsymbol{\Psi}}}%
\global\long\def\hatbolOmega{\widehat{\boldsymbol{\Omega}}}%
\global\long\def\barbolmu{\overline{\bolmu}}%
\global\long\def\barmbX{\overline{\mbX}}%
\global\long\def\mbbR{\mathbb{R}}%
\global\long\def\mbbP{\mathbb{P}}%
\global\long\def\mbbQ{\mathbb{Q}}%
\global\long\def\mbbS{\mathbb{S}}%
\global\long\def\mbbH{\mathbb{H}}%
\global\long\def\mbbX{\mathbb{X}}%
\global\long\def\mbbY{\mathbb{Y}}%
\global\long\def\mbbG{\mathbb{G}}%
\global\long\def\mbbZ{\mathbb{Z}}%
\global\long\def\spc{\mathcal{S}}%
\global\long\def\calA{\mathcal{A}}%
\global\long\def\calB{\mathcal{B}}%
\global\long\def\calC{\mathcal{C}}%
\global\long\def\calD{\mathcal{D}}%
\global\long\def\calE{\mathcal{E}}%
\global\long\def\calF{\mathcal{F}}%
\global\long\def\calG{\mathcal{G}}%
\global\long\def\calH{\mathcal{H}}%
\global\long\def\calI{\mathcal{I}}%
\global\long\def\calJ{\mathcal{J}}%
\global\long\def\calK{\mathcal{K}}%
\global\long\def\calL{\mathcal{L}}%
\global\long\def\calM{\mathcal{M}}%
\global\long\def\calN{\mathcal{N}}%
\global\long\def\calO{\mathcal{O}}%
\global\long\def\calP{\mathcal{P}}%
\global\long\def\calQ{\mathcal{Q}}%
\global\long\def\calR{\mathcal{R}}%
\global\long\def\calS{\mathcal{S}}%
\global\long\def\calT{\mathcal{T}}%
\global\long\def\calU{\mathcal{U}}%
\global\long\def\calV{\mathcal{V}}%
\global\long\def\calW{\mathcal{W}}%
\global\long\def\mbell{\boldsymbol{\ell}}%
\global\long\def\bolell{\boldsymbol{\ell}}%
\global\long\def\mbzero{\mathbf{0}}%
\global\long\def\bolPhio{\boldsymbol{\Phi}_{0}}%
\global\long\def\bolOmegao{\boldsymbol{\Omega}_{0}}%
\global\long\def\bolSigmaX{\bolSigma_{\mbX}}%
\global\long\def\bolSigmaY{\bolSigma_{\mbY}}%
\global\long\def\bolSigmaXY{\boldsymbol{\Sigma}_{\mbX\mbY}}%
\global\long\def\mbSX{\mathbf{S}_{\mbX}}%
\global\long\def\mbSY{\mathbf{S}_{\mbY}}%
\global\long\def\mbSXY{\mathbf{S}_{\mbX\mbY}}%
\global\long\def\mbSYX{\mathbf{S}_{\mbY\mbX}}%
\global\long\def\mbRYX{\mathbf{S}_{\mbY|\mbX}}%
\global\long\def\mbRXY{\mathbf{S}_{\mbX|\mbY}}%
\global\long\def\mbSc{\mbS_{\mbC}}%
\global\long\def\mbSd{\mbS_{\mbD}}%
\global\long\def\tilmbX{\widetilde{\mbX}}%
\global\long\def\tilmbZ{\widetilde{\mbZ}}%
\global\long\def\tilmbS{\widetilde{\mbS}}%
\global\long\def\tilmbY{\widetilde{\mbY}}%
\global\long\def\Xbar{\overline{\mbX}}%
\global\long\def\Ybar{\overline{\mbY}}%
\global\long\def\mbSm{\mathbf{S}_{-}}%
\global\long\def\mbSym{\mathbf{S}_{\mbY-}}%
\global\long\def\tilbolPhi{\widetilde{\bolPhi}}%
\global\long\def\tilbolPsi{\widetilde{\boldsymbol{\Psi}}}%
\global\long\def\sumn{\sum_{i=1}^{n}}%
\global\long\def\E{\mathrm{E}}%
\global\long\def\F{\mathrm{F}}%
\global\long\def\J{\mathrm{J}}%
\global\long\def\H{\mathrm{H}}%
\global\long\def\G{\mathrm{G}}%
\global\long\def\Cov{\mathrm{cov}}%
\global\long\def\cov{\mathrm{cov}}%
\global\long\def\Corr{\mathrm{corr}}%
\global\long\def\Var{\mathrm{var}}%
\global\long\def\dimension{\mathrm{dim}}%
\global\long\def\spn{\mathrm{span}}%
\global\long\def\vech{\mathrm{vech}}%
\global\long\def\vecc{\mathrm{vec}}%
\global\long\def\Prob{\mathrm{Pr}}%
\global\long\def\Env{\mathrm{env}}%
\global\long\def\tr{\mathrm{tr}}%
\global\long\def\dg{\mathrm{diag}}%
\global\long\def\asyVar{\mathrm{avar}}%
\global\long\def\MSE{\mathrm{MSE}}%
\global\long\def\OLS{\mathrm{OLS}}%
\global\long\def\CS{\calS_{Y\mid\mbX}}%
\global\long\def\covenv{\calE_{\bolDelta_{0}}(\bolbeta)}%
\global\long\def\tilbolbeta{\widetilde{\boldsymbol{\beta}}}%
\global\long\def\tilbolSigma{\widetilde{\bolSigma}}%
\global\long\def\mbbE{\mathbb{E}}%
\global\long\def\khat{\widehat{k}}%
\global\long\def\lhat{\widehat{l}}%
\global\long\def\hattau{\widehat{\tau}}%
\global\long\def\hatsigmaj{\widehat{\sigma}_{j}^{2}}%
\global\long\def\hatsigmanj{\widehat{\sigma}_{nj}^{2}}%
\global\long\def\hatD{\widehat{D}}%
\global\long\def\tildelta{\widetilde{\delta}}%

\newcommand{\change}[1]{{\leavevmode\color{red}{#1}}}

\title{{Significance testing for canonical correlation \\
		analysis in high dimensions}}
\author{
\bigskip
Ian W.~McKeague$^\dag$ and Xin Zhang$^\ddag$\\
\normalsize{\textit{$^\dag$Columbia University and $^\ddag$Florida State University}}
}

\maketitle
\begin{abstract}
	We consider the problem of testing for the presence of 
	linear relationships between large sets of random variables based on a post-selection inference approach to canonical correlation
	analysis. The challenge is to adjust for the selection of subsets
	of variables having linear combinations with maximal sample correlation.
	To this end, we construct a stabilized one-step estimator of the euclidean-norm of the canonical correlations 
	maximized
	over subsets of variables of pre-specified cardinality.
	This estimator is shown to be consistent for its target parameter and asymptotically normal provided the dimensions
	of the variables do not grow too quickly with sample size. We also develop
	a greedy search algorithm to accurately compute the estimator,
	leading to a computationally tractable omnibus test for the global
	null hypothesis that there are no linear relationships between any
	subsets of variables having the pre-specified cardinality. Further,
	we develop a confidence interval for the target parameter that takes the variable selection into account. \\
	
\end{abstract}


\section{Introduction}

When exploring the relationships between two sets of variables measured
on the same set of observations, canonical correlation analysis \citep[CCA;][]{hotelling1936relations}
sequentially extracts linear combinations with maximal sample correlation.
Specifically, with $\mbX\in\mbbR^{p}$ and $\mbY\in\mbbR^{q}$ as
two random vectors,  the first step of CCA  targets the parameter
\begin{equation}
\rho=\max_{\bolalpha\in\mbbR^{q},\bolbeta\in\mbbR^{p}}\mathrm{corr}(\bolalpha^{T}\mbY,\bolbeta^{T}\mbX),\quad\mathrm{subject\ to}\ \Var(\bolalpha^{T}\mbY)=1=\Var(\bolbeta^{T}\mbX).\label{CCA}
\end{equation}
Subsequent steps of CCA repeat this process subject to the constraint
that the next linear combinations of $\mbX$ (and $\mbY$) are uncorrelated with earlier ones, giving
a decreasing sequence of correlation coefficients. We are interested
in testing whether the maximal canonical correlation coefficient $\rho\neq0$
versus the null hypothesis $\rho=0$ in the high-dimensional setting
in which $p$ and $q$ grow with sample size $n$. This is equivalent to
testing whether all of the canonical correlation coefficients vanish, or whether their sum of squares $\tau^{2}$ (known as  the \citet{pillai1955} trace) vanishes.

Over the last dozen years, numerous sparse canonical correlation analysis
(SCCA) methods \citep[e.g.,][]{witten2009penalized,hardoon2011sparse, gao2017sparse,mai2019iterative,Qadar2019,shu2020}
have been developed as extensions of classical CCA by adapting  regularization
approaches from regression, e.g., lasso \citep{lasso}, elastic net \citep{elasticnet} and soft thresholding.  

SCCA methods have been widely applied to high-dimensional omics data to detect associations between gene expression and DNA copy number/polymorphisms/methylation, with the aim of revealing networks of co-expressed and co-regulated genes \citep{waa2007,waa2008, naylor2010, parkhomenko2009sparse, wang2015inferring}. A problem with the indiscriminate use of such methods, however, is \textit{selection bias}, arising
when the effects of variable selection on subsequent statistical analyses
are ignored, i.e., failure to take into account ``double dipping''
of the data when assessing evidence of association.

Devising valid tests for associations in high-dimensional SCCA, along
with confidence interval estimation for the strength of the association,
poses a challenging post-selection inference problem. Nevertheless,
some progress on this problem has been made. \citet{yang2015} proposed
the sum of  sample canonical correlation coefficients as
a test statistic and established a valid calibration under the sparsity assumption that the number of non-zero canonical correlations is finite and fixed, with the dimensions $p$ and $q$  proportional to sample size. Their approach comes at the cost of assuming that $\mbX$
and $\mbY$ are  jointly Gaussian (and thus fully independent under the null); similar results for the maximal  sample canonical correlation coefficient are developed in 
\citet{bao2019}.  
\citet{zheng2019} developed a test for the presence of correlations
among arbitrary components of a given high-dimensional random vector,
for both sparse and dense alternatives, but their approach also requires
an independent components structure.

In this paper, we provide valid post-selection inference for a new
version of SCCA in high-dimensional settings. We obtain a computationally tractable and asymptotically valid confidence interval for 
$\tau_{{\max}}$, where $\tau_{{\max}}^2$ is the maximum of the Pillai trace over all subvectors
of $\mbX$ and $\mbY$ having prespecified dimensions $s_{x}$ and
$s_{y}$, respectively. The method is fully nonparametric in the sense
that no distributional assumptions or sparsity assumptions are required. Rather than adopting
a penalization approach or making a sparsity assumption on the number of non-zero canonical correlations to regularize the problem, we use the sparsity
levels $s_{x}\ll p$ and $s_{y}\ll q$ for regularization, and also
for controlling the computational cost of searching over large collections
of subvectors. We introduce a test statistic $\widehat\tau_{{\max}}$ constructed
as a stabilized and efficient one-step estimator of $\tau_{{\max}}$.
Then, assuming $p$ and $q$ do not grow too quickly with sample size, specifically that {$\log(p+q)/\sqrt n \to 0$}, 
we show that a studentized version of $\widehat\tau_{{\max}}$ (after centering
by $\tau_{{\rm max}}$) converges weakly to standard normal. This leads to a practical way of calibrating a formal omnibus test for the global
	null hypothesis ($\tau_{{\rm max}}=0$) that there are no linear relationships between any
	subsets of variables having the pre-specified cardinality, along with an asymptotically valid Wald-type confidence interval
for $\tau_{{\rm max}}$. 

The proposed approach applies to any choice of pre-specified sparsity levels $s_x$ and $s_y$, which do not need to be the same as the true number of ``active'' variables 
in the population CCA, although they should be sufficiently large to capture the key associations.  
The test procedure and confidence interval for the target parameter $\tau_{{\max}}$ are  asymptotically valid for any pre-specified sparsity levels, and work well provided the  sample cross-covariance matrices between subvectors of  $\mbX$
and $\mbY$ having dimensions $s_x$ and $ s_y$ are  sufficiently accurate.

Our approach is related to the type of post-selection inference procedure
for marginal screening developed by \citet{mckeague2015adaptive},
which applies to the one-dimensional response case ($q=1$ in the
present notation). To extend this approach to the SCCA setting, in
which both $p$ and $q$ can be large, requires a trade-off between
computational tractability and statistical power. The calibration
used in \citet{mckeague2015adaptive} is a double-bootstrap technique,
which is computationally expensive. To obtain a fast calibration
method for SCCA, we adapt the sample-splitting stabilization technique
of \citet{luedtke2018parametric} to the SCCA setting, which provides calibration using
a standard normal limit. Further, to control the computational complexity of searching
through large collections of subvectors of $\mbX$ and $\mbY$ when computing $\widehat\tau_{{\max}}$, we
develop a greedy search algorithm related to that of \citet{wiesel2008greedy}.

The rest of the article is organized as follows. Section~\ref{sec:method} introduces the population target parameter $\tau_{{\max}}$ and develops its stabilized one-step  estimator, taking the non-regularity of $\tau_{{\max}}$ at the global null hypothesis into account; asymptotic results are given in Section~\ref{sec:theory}. Section~\ref{sec:algorithm} proposes the greedy search algorithm to speed up the computation and provides a rationale based on submodularity.  Sections~\ref{sec:sim} and \ref{sec:real} respectively contain a simulation study and a real data example using data collected under the Cancer Genome Atlas Program \citep{weinstein2013cancer}.  Section~\ref{sec:discussion} concludes the paper with a short discussion. The Appendix  contains  a derivation of the influence function of the Pillai trace, which plays a key role in its efficient estimation, and the proof of an identity involving increments of the Pillai trace used in the greedy search algorithm. The Supplementary Materials collect all additional technical details, numerical results and  \textsf{R}  code. 

\section{Test procedure\label{sec:method}}

\subsection{Preliminaries}

Let $\bolSigma_{\mbX}>0$ and $\bolSigma_{\mbY}>0$ denote the (invertible) 
covariance matrices of $\mbX$ and $\mbY$, with cross-covariance
matrix $\bolSigma_{\mbX\mbY}$ and   standardized cross-covariance  matrix $\bolLambda_{\mbX\mbY}\equiv\bolSigma_{\mbX}^{-1/2}\bolSigma_{\mbX\mbY}\bolSigma_{\mbY}^{-1/2}\in\mbbR^{p\times q}$ (also known as the coherence matrix). The sample counterparts
are denoted $\mbS_{\mbX}$, $\mbS_{\mbY}$,
$\mbS_{\mbX\mbY}$ and $\mbC_{\mbX\mbY}$, respectively.

The coherence matrix $\bolLambda_{\mbX\mbY}$ has   $\min(p,q)$ singular values; when  listed in decreasing order they coincide with the canonical correlation coefficients,  and $\rho$ defined in \eqref{CCA} is the largest.  A closely related parameter 
in MANOVA is the Pillai trace  $\tau^2$ \citep{pillai1955}, defined as the sum of squares of the canonical correlation coefficients, or  equivalently  
\begin{equation}
\tau^{2}=\Vert \bolLambda_{\mbX\mbY} \Vert_{F}^2 = \tr(\bolLambda_{\mbX\mbY}\bolLambda_{\mbX\mbY}^{T})=\tr\{\mbH(\mbH+\mbE)^{-1}\},
\end{equation}
where  $\Vert\cdot\Vert_{F}$ is  Frobenius
norm, and $\mbH=\bolSigma_{\mbY\mbX}\bolSigma_{\mbX}^{-1}\bolSigma_{\mbX\mbY}$ and $\mbE=\bolSigma_{\mbY}-\mbH$ are
  population versions of covariance matrices in  a linear model for predicting $\mbY$ from $\mbX$. Specifically,  $\mbH$ 
is the covariance matrix of the least-squares-predicted outcome 
in the linear model $\mbY=\mbA+\mbB\mbX+\bolvarepsilon$,
where $\Cov(\bolvarepsilon)=\mbE$ and $\bolvarepsilon$ is uncorrelated with $\mbX$. 

We will need some general concepts from semi-parametric efficiency theory. Suppose we observe a general random vector $\mbO\sim P$.
Let $L_0^2(P)$ denote the Hilbert space of $P$-square integrable functions with mean zero. Consider a smooth one-dimensional family of probability
measures $\{ P_t, t \in [0,1]\} $ with $P_0=P$ and having score function $k\in L_0^2(P)$ at $t=0$. The tangent space $T(P)$ is the
$L_0^2(P)$-closure of the linear span of all such score functions $k$. For example, if nothing is known about $P$, then
$P_t(d\mbo)=(1+t k(\mbo))P(d\mbo)$ is such a submodel for any bounded function $k$ with mean zero (provided $t$ is sufficiently small),
so $T(P)$ is seen to be the whole of $L_0^2(P)$ in this case. 
Let $\psi(P)$ be a real parameter that is pathwise differentiable at $P \colon$ there exists
$g\in L_0^2(P)$ such that $\lim_{t\to 0} \left ( \psi(P_t)-\psi(P)\right)/t =\langle g,k\rangle$, for any smooth submodel
$\{ P_t\}$ with score function $k$, where $\langle \cdot,\cdot\rangle$ is the inner product in $L_0^2(P)$. The function $g$ is
called a gradient (or influence function) for $\psi$; the projection $\IF_\psi$ 
of any gradient into the tangent space $T(P)$ is unique and is known as
the canonical gradient (or efficient influence function).  The supremum of the Cram\'er--Rao bounds for all submodels (the information bound)
is given by the second moment of $\IF_\psi(\mbO)$.  Furthermore, the influence function as  derived using  von Mises calculus \citep[Chapter 20] {Vaart2000} of any regular and asymptotically linear estimator must be a gradient \citep[Proposition 2.3]{Pfanzagl1990}.

A one-step estimator is an empirical bias correction of a na\"{i}ve plug-in estimator in the direction of a gradient of the parameter of interest \citep{Pfanzagl1982}; when this gradient is the canonical gradient, then this results in an efficient estimator under some regularity conditions. Given an initial estimator $\widehat{P}$ of $P$ and any gradient $D(\widehat{P})$ of the parameter $\psi$ evaluated at $\widehat{P}$,  we have
$  \psi(\widehat{P}) - \psi(P)= -P D(\widehat{P}) + {\rm Rem}_{\psi}(\widehat{P},P)$,
where ${\rm Rem}_{\psi}(\widehat{P},P)$ is negligible if $\widehat{P}$ is close to $P$ in an appropriate sense. Here $Pf $ denotes the expectation under $P$ of a random real-valued function $f$, ignoring the randomness in $f$.  As $D(P)$ has mean zero under $P$, we expect that $P D(\widehat{P})$ is close to zero if $D$ is continuous in its argument and $\widehat{P}$ is close to $P$. However, the rate of convergence of $P D(\widehat{P})$ to zero as sample size grows may be slower than $n^{-1/2}$. The one-step estimator aims to improve $\psi(\widehat{P})$ and achieve $n^{1/2}$-consistency and asymptotically normality by adding an empirical estimate $\mathbb{P}_n D(\widehat{P})$ of its deviation from $\psi(P)$. The one-step estimator $\widehat{\psi}\equiv \psi(\widehat{P}) + \mathbb{P}_n D(\widehat{P})$ then satisfies the expansion
$
  \widehat{\psi} - \psi(P)= (\mathbb{P}_n-P) D(\widehat{P}) + {\rm Rem}_{\psi}(\widehat{P},P)$.
Under an empirical process and $L^2(P)$ consistency condition on $D(\widehat{P})$, the leading term on the right is asymptotically equivalent to $(\mathbb{P}_n-P) D(P)$, which converges in distribution to a mean-zero Gaussian 
limit with consistently estimable covariance. To minimize the variance of the Gaussian limit,  $D(\widehat{P})$ can be taken as the canonical gradient of $\psi$ at $\widehat{P}$.  


\subsection{Maximal Pillai trace}
\label{target}

Clearly, the null hypotheses $\rho=0$ and $\tau=0$ are 
equivalent, but the root-Pillai trace $\tau$ (the positive square root of the Pillai trace) is a more informative target parameter  than the leading canonical correlation $\rho$, although the two would coincide if there is  only a single non-zero canonical correlation coefficient. Moreover, because estimating the maximal values of  $\tau$ or $\rho$, subject to sparsity constraints,   needs repeated evaluation  and updating of the estimates, the choice of $\tau$ provides considerable  computational savings over $\rho$, as the latter would require updating the entire eigen-decomposition at each step.  

Our approach is to develop asymptotic distribution results for a regularized empirical version of this target parameter when the dimensions $p$ and $q$ grow with sample size $n$. Specifically, given sparsity levels $s_{x}$ and $s_{y}$ for $\mbX$ and $\mbY$,
respectively,  we are interested in selecting  index sets $\calK\subset\{1,\dots,p\}$
and $\calJ\subset\{1,\dots,q\}$ with cardinality $\vert\calK\vert\leq s_{x}$
and $\vert\calJ\vert\leq s_{y}$ that maximize the Pillai trace (or equivalently the root-Pillai trace) of their corresponding sub-vectors. The sparsity levels $s_x$ and $s_y$ are pre-specified and fixed, e.g., $(s_{x},s_{y})=(1,2)$. 

Given independent observations $\mbO_{i}=(\mbX_{i}^{T},\mbY_{i}^{T})^{T}$,
$i=1,\dots,n$, drawn from a distribution $P$ on $\mbbR^{p+q}$, 
we target the non-regular
parameter  
\begin{equation}
\label{maxtau}
\tau_{\rm max}\equiv\max_{d\in\calD_{n}}\Psi^{d}(P),
\end{equation}
where   
$\calD_{n}=\left\{ (\calJ,\calK)\mid\vert\calK\vert=s_{x}\le p,\vert\calJ\vert=s_{y}\le q,\calK\subseteq\{1,\dots,p\},\calJ\subseteq\{1,\dots,q\}\right\}$ and $\Psi^{d}(P)=\Vert\bolLambda_{\mbX_{\calK}\mbY_{\calJ}}\Vert_{F}$.
Note that $\tau_{\rm max}\le \tau$ with equality when $s_x=p$, $s_y=q$. The subscript $n$ in $\calD_{n}$ indicates that the dimensions
$p=p_n$ and $q=q_n$ are allowed to increase with $n$.  

The numbers of active variables  $(s_x^\star,s_y^\star)$ are the smallest values of the sparsity levels $(s_x, s_y)$ for which $\tau_{\rm max}= \tau$. Note that  $s_x^\star$ and $s_y^\star$ can be as large as $p$ and $q$, respectively, and as small as the number of non-zero canonical correlation coefficients for $\mbX$ and $\mbY$ (the rank of $\bolLambda_{\mbX\mbY}$, denoted $K$ in the sequel). The non-regularity   arises for various reasons, including the fact that multiple
elements of $\calD_{n}$ may achieve the same maximum in (\ref{maxtau}) (e.g., when $\tau_{\rm max}=0$). This may occur, for example, if the pre-specified sparsity levels are  larger than the true sparsity levels  $(s_x^\star,s_y^\star)$, but as we see later in this section the sample root-Pillai trace is  non-regular at  $\tau_{\rm max}=0$ even when $d$ is fixed.  



We  now use von Mises calculus to derive the canonical gradient $D^{d}(P)(\mbo)$
of  the functional $\Psi^{d}(P)$ for a fixed 
$d\in \calD_{n}$.  This canonical gradient      
can be found in terms of the influence function of its square  $\Phi^d (P) = \{\Psi^{d}(P)\}^2$, and using the fact that the tangent space is the whole of $L_0^2(P)$ in this nonparametric setting.  Let  $P_{\epsilon}=(1-\epsilon)P+\epsilon\delta_{\mbo}$, where
$\epsilon\in [0,1]$ and $\delta_{\mbo}$ is the Dirac
measure at the point $\mbo=(\mbx^{T},\mby^{T})^{T}$.  When $\Psi^{d}(P)>0$, we have
\begin{eqnarray}
D^{d}(P)(\mbo) & = & \left.\dfrac{d\Psi^{d}(P_{\epsilon})}{d\epsilon}\right|_{\epsilon=0}
= {1\over 2\Psi^d (P)}  \left.\dfrac{d\Phi^{d}(P_{\epsilon})}{d\epsilon}\right|_{\epsilon=0} \label{can1} \end{eqnarray}
where 
\begin{eqnarray}\left.\dfrac{d\Phi^{d}(P_{\epsilon})}{d\epsilon}\right|_{\epsilon=0}& = & -\{\mbx_{\calK}-\E_{P}(\mbX_{\calK})\}^{T}\bolSigma_{\mbX_{\calK}}^{-1}\bolSigma_{\mbX_{\calK}\mbY_{\calJ}}\bolSigma_{\mbY_{\calJ}}^{-1}\bolSigma_{\mbY_{\calJ}\mbX_{\calK}}\bolSigma_{\mbX_{\calK}}^{-1}\{\mbx_{\calK}-\E_{P}(\mbX_{\calK})\}\nonumber \\
& & - \{\mby_{\calJ}-\E_{P}(\mbY_{\calJ})\}^{T}\bolSigma_{\mbY_{\calJ}}^{-1}\bolSigma_{\mbY_{\calJ}\mbX_{\calK}}\bolSigma_{\mbX_{\calK}}^{-1}\bolSigma_{\mbX_{\calK}\mbY_{\calJ}}\bolSigma_{\mbY_{\calJ}}^{-1}\{\mby_{\calJ}-\E_{P}(\mbY_{\calJ})\}\nonumber \\
&  & +2\{\mby_{\calJ}-\E_{P}(\mbY_{\calJ})\}^{T}\bolSigma_{\mbY_{\calJ}}^{-1}\bolSigma_{\mbY_{\calJ}\mbX_{\calK}}\bolSigma_{\mbX_{\calK}}^{-1}\{\mbx_{\calK}-\E_{P}(\mbX_{\calK})\}.\label{CanonicalGradient}
\end{eqnarray}
The details are given in  Appendix A.1, where we also show that $\E_{P}\{D^{d}(P)(\mbO)\}=0$, so the influence function belongs to the tangent space $L_0^2(P)$ and is  thus the efficient influence function.   

A continuous extension of  $D^{d}(P)(\mbo) $ to the case $\Psi^d(P)=0$ is obtained as follows.  The matrix-valued parameter  $\psi(P)=\bolLambda \equiv \bolLambda_{\mbY_{\calJ}\mbX_{\calK}}$  is pathwise differentiable, so when $\psi(P)=0$ there exists a matrix ${\bf G}$ (which we can take as the efficient  influence function) of the same dimensions as $\bolLambda$ and having entries in $L_0^2(P)$ such that 
$\psi(P_t)/t \to \langle {\bf G}, k\rangle$ as $t\to 0$
for any smooth one-dimensional parametric sub-model $\{P_t, t \in [0,1]\}$ with score function $k\in L_0^2(P)$ at $t =0$.  Here the inner product notation in $\langle {\bf G}, k\rangle$  
 is understood to be applied  entry-wise to ${\bf G}$. 
Writing $\bolLambda_{t} \equiv \psi(P_t)$, and arranging that it does not vanish at any $t$ apart from $t=0$,
it follows that 
$\bolLambda_{t}/\Vert\bolLambda_{t}\Vert_{F}\to \langle {\bf G}, k\rangle/ \Vert \langle {\bf G}, k\rangle\Vert_{F} \equiv  {\bf L}$ in Frobenius norm as $t\to 0$.  It is then easily checked that for each fixed $\mbo$, 
\begin{equation}
D^{d}(P)(\mbo) \equiv \lim_{t \to 0}D^{d}(P_t)(\mbo) = \{\mby_{\calJ}-\E_{P}(\mbY_{\calJ})\}^{T}\bolSigma_{\mbY_{\calJ}}^{-1/2}{\bf L}\bolSigma_{\mbX_{\calK}}^{-1/2}\{\mbx_{\calK}-\E_{P}(\mbX_{\calK})\},\label{newDefCG}
\end{equation}
providing the  canonical gradient of $\Psi^{d}(P)$ when $\Psi^d(P)=0$.  

For univariate $X$ and $Y$, the functional  $P\mapsto \Corr_P(X,Y)$ has canonical gradient 
$$
 \dfrac{\{x-\E_P(X)\}\{y-\E_P(Y)\}}{\sqrt{\Var(X)\Var(Y)}} 
	 - \dfrac{\Corr(X,Y)}{2\Var(X)}\{x-\E_P(X)\}^{2}-\dfrac{\Corr(X,Y)}{2\Var(Y)}\{y-\E_P(Y)\}^{2},
$$
 a result due to
Colin Mallows \citep{Devlin1975}.
When $\Corr(X,Y)=0$ the last two terms above drop out, and  the expression agrees with the canonical gradient of $\Psi^{d}(P)$ in  (\ref{newDefCG}), since  ${\bf L}=1$ in this case. 
In the multivariate case, the entries of the matrix ${\bf L}$ are nuisance  parameters that are absent in the univariate case. 

The nuisance parameters in $\bf L$ vary with $d$ and the score function $k$, indicating the presence of non-regularity in the root-Pillai trace at zero, as the underlying $k$ is not identifiable (it  plays the role of a local parameter).  When target parameters  take values on the boundary 
of their parameter space (zero is on the boundary  in our case),  non-regularity is known to cause unstable asymptotics, such as  inconsistency of the bootstrap, even in the simple example of a population mean restricted to be non-negative \citep{andrews2000}.  That is, dependence of a canonical gradient (or efficient influence function)  on an arbitrary score function implies unstable behavior of the estimator, especially in small samples. 
This form of non-regularity  is present  in dimensions $p\ge 2$ and $q\ge 2$ (even without selection of  $d\in \calD_{n}$), but not in the case of  univariate $X$ and $Y$  since the parameter space for the  correlation coefficient is taken as the open interval $(-1,1)$, which has no boundary.  

This boundary type of non-regularity is distinct from the post-selection type of non-regularity  noted  by  \citet[Section 2]{mckeague2015adaptive} in the case  $p\ge 2$ and $q=1$, in which the asymptotic distribution of the maximal absolute sample correlation is discontinuous  at $\tau_{\rm max}= 0$.  This type of non-regularity  occurs in the present setting  with the   sample estimator of $\tau_{\rm max}$ given by
\begin{equation}
\widehat{\tau}_{\rm samp}=\max_{\vert\calK\vert\leq s_{x},\vert\calJ\vert\leq s_{y}}\Vert\mbC_{\mbY_{\calJ}\mbX_{\calK}}\Vert_{F}=\max_{\vert\calK\vert=s_{x},\vert\calJ\vert=s_{y}}\Vert\mbC_{\mbY_{\calJ}\mbX_{\calK}}\Vert_{F},\label{eq:cca_screening}
\end{equation}
where $\mbY_{\calJ}\in\mbbR^{s_{y}}$ and $\mbX_{\calK}\in\mbbR^{s_{x}}$
are the selected variables.  Here the second equality is a  direct consequence of Lemma~\ref{lem: stepwise} in the sequel.  It is challenging to use the estimator $\widehat{\tau}_{\rm samp}$ as a test statistic for the global null hypothesis that $\tau_{\rm max}=0$ because of the discontinuity in its asymptotic distribution at the null, but the stabilized one-step estimator $\widehat{\tau}_{\rm max}$ introduced below avoids this difficulty.

Curiously, the boundary-type of non-regularity does not arise with the Pillai trace itself, since its canonical gradient \eqref{CanonicalGradient} does not depend on any  score function $k$; an intuitive explanation is that by squaring the root-Pillai trace, the non-regularity is smoothed out at zero.  However, this squaring has the effect of causing severe bias in the sampling distribution of the stabilized one-step estimator of $\tau_{\rm max}^2$, especially when  $\tau_{\rm max}$ is small and in small samples (see  Figures \ref{fig:null} and \ref{fig:alt} in Section \ref{hyp-sim}).  This problem does not arise with $\widehat\tau_{\rm max}$, hence our focus in the sequel on the root-Pillai trace.

Many authors have studied hypothesis testing  problems in which a nuisance parameter is only identifiable under the alternative \citep[e.g.,][]{Davies1977,Davies1987,Davies2002,hansen1996inference}.  
Here we encounter the situation where nuisance parameters appear only in the null, so calibration of the test statistic may potentially depend on $\mbL$.   \cite{Leeb2017} have studied a post-selection calibration  method that uses estimates of  such nuisance parameters, but, as we will see,  our approach  leads to an asymptotically pivotal estimator of $\tau_{\max}$ without the need to estimate $\mbL$.


\subsection{Stabilized one-step estimator\label{sec:estimation}}
In this section we develop  the stabilized one-step estimator  for the target parameter $\tau_{\max}$ 
in terms of the canonical gradient $D^{d}(P)$, which will be estimated by plugging-in  empirical distributions  in place of $P$ in  \eqref{can1}.  
The data are first randomly ordered and we  consider subsamples consisting of the first $j$ observations for $j=\ell_n,\ldots,n-1$, where $\{\ell_n\}$ is some positive integer sequence such that both $\ell_n$ and $n-\ell_n$ tend to infinity.  In practice, we recommend randomly ordering the data  say $K=10$ times, and then combining the $K$ confidence intervals by averaging (for more details see the real data example). 
Let $P_{j}$ be the empirical distribution
of the first $j$ observations.  
The following procedure is a version of the construction of the stabilized one-step estimator in \citet{luedtke2018parametric}.

For each $j=\ell_{n},\dots,n-1$, 
compute the following quantities:
\begin{enumerate}
	\item The selected subsets of variables $d_{nj}=(\widehat{\calK},\widehat{\calJ})$ given by
	\begin{equation}\label{update_dnj}
	d_{nj}\equiv\argmax_{d\in\calD_{n}}\Psi^{d}(P_{j})=\argmax_{\vert\calK\vert=s_x,\ \vert\calJ\vert=s_y}\Vert\mbC_{\mbX_{\calK}\mbY_{\calJ}}(P_{j})\Vert_{F}.
	\end{equation}
	\item The corresponding maximum $\Psi^{d_{nj}}(P_{j})=\Vert\mbC_{\mbX_{\widehat{\calK}}\mbY_{\widehat{\calJ}}}(P_{j})\Vert_{F}$
	and $\widehat{D}_{j}(\mbO_{j+1})\equiv D^{d_{nj}}(P_{j})(\mbO_{j+1})$
	using the canonical gradient given by \eqref{can1} and \eqref{CanonicalGradient} with $P=P_j$. 
	\item An estimate of the variance of $\widehat{D}_{j}(\mbO_{j+1})$:
	$$
	\hatsigmaj=\dfrac{1}{j}\sum_{i=1}^{j}\left\{ \hatD_{j}(\mbO_{i})-\dfrac{1}{j}\sum_{m=1}^{j}\hatD_{j}(\mbO_{m})\right\} ^{2}.
	$$
	\item Weights $w_{j}=\overline{\sigma}_{n}/\widehat{\sigma}_{j}$, where $\overline{\sigma}_{n}=\left(\dfrac{1}{n-\ell_{n}}\sum_{j=\ell_{n}}^{n-1}\widehat{\sigma}_{j}^{-1}\right)^{-1}$ is the harmonic mean of the $\widehat{\sigma}_{j}$, $j=\ell_{n},\dots,n-1$. 
\end{enumerate}
The stabilized one-step estimator for the target parameter  $\tau_{{\rm max}}$ is then given by 
	\begin{equation}\label{taumax_hat}
\widehat\tau_{\max}=\dfrac{1}{n-\ell_{n}}\sum_{j=\ell_{n}}^{n-1}w_{j}\left\{ \Psi^{d_{nj}}(P_{j})+\hatD_{j}(\mbO_{j+1})\right\} ,
	\end{equation}
and an asymptotic $100(1-\alpha)$\% Wald-type confidence interval for $\tau_{\max}$
is 
\begin{equation}
[\mathrm{LB}_{n},\mathrm{UB}_{n}]=\left[\widehat\tau_{\max}-z_{\alpha/2}\dfrac{\overline{\sigma}_{n}}{\sqrt{n-\ell_{n}}},\widehat\tau_{\max}+z_{\alpha/2}\dfrac{\overline{\sigma}_{n}}{\sqrt{n-\ell_{n}}}\right],
\end{equation}
where $z_{\alpha/2}$ is the upper $\alpha/2$-quantile of standard normal. For an $\alpha$-level test of $\tau_{\max}=0$ versus $\tau_{\max}>0$,  reject the null hypothesis $\tau_{\max}=0$ if the lower bound of the $100(1-2\alpha)\%$ confidence interval exceeds 0.  The estimator $\widehat\tau_{\max}$ is a weighted version of the ``online''  one-step estimator introduced by \cite{laan2014}, where in our case $\Psi^{d_{nj}}(P_{j})$ is improved using its estimated canonical gradient evaluated at a new observation.   

Recursive properties of the algorithm allow considerable speed-up in the computation (see Section \ref{sec:recur} of the Appendix).
Further, when the sample size $n$ is large, we follow \citet{luedtke2018parametric}'s suggestion of  speeding up the  $(n-\ell_n)$ updates by restricting  the sample stream over $j=\ell_n,\dots,n-1$ to only involve increments in $j$ of size $C\ge 2$. 
The asymptotic properties of the stabilized one-step estimator are not affected by $C$. In our experience, the  results are insensitive to the choice of $C$, provided  $n$ is sufficiently large relative to $C$.  We fixed $C=20$ and $\ell_n= \lceil n/2\rceil$ in our numerical studies.  Methods   for estimating the number of non-zero canonical correlation coefficients $K$ have been extensively studied in the signal processing literature \citep[e.g.,][]{song2016,Seghouane2019}, and these can  be used to provide a lower bound on the choice of $s_x$ and $s_y$.  In practive, however, we recommend specifying $s_x$ and $s_y$ via a graphical inspection of the increments in the sample Pillai trace as the sparsity levels  increase, see  Figure \ref{fig:forwardstep} in Section \ref{sec:algorithm}.

\subsection{Asymptotic  results \label{sec:theory} }


We assume that each variable $X_{k}$, $k=1,\dots,p$,
or $Y_{j}$, $j=1,\dots,q,$ is bounded within $[-1,1]$, and that
the canonical gradient of $\Psi^d(P)$ satisfies 
\begin{equation}
\inf_{n\geq2}\min_{d\in\calD_{n}}\Var_{P}\{D^{d}(P)(\mbO)\}\geq\gamma,\label{C2}
\end{equation}
for some constant $\gamma>0$. This condition is  mild in view of  \eqref{can1} and  \eqref{newDefCG}, and  is imposed to ensure a non-degenerate asymptotic distribution for the one-step estimator, as needed to form non-trivial confidence intervals for $\tau_{\rm max}$. 
To ensure that the canonical gradient is uniformly bounded
for all $d$, we also assume that, for some $\delta>0$, 
\begin{equation}
\sup_{d\in\calD_{n}}\max\{\Vert\bolSigma_{\mbX_{\calK}}^{-1}\Vert,\Vert\bolSigma_{\mbY_{\calJ}}^{-1}\Vert\}<\delta^{-1},\label{C1}
\end{equation}
where $\Vert\mbM\Vert$ denotes the largest singular value of matrix
$\mbM$ (or the operator norm). This mild condition
means that the smallest eigenvalue of $\bolSigma_{\mbX_{\calK}}$
(or $\bolSigma_{\mbY_{\calJ}}$) is greater than some constant $\delta$. 
We treat $\delta$, $\gamma$ and $s=s_{x}=s_{y}$
as fixed, and thus omit the dependence
on $\delta$, $\gamma$ and $s$ in the asymptotic statements. On
the other hand, we allow both  dimensions $p$ and $q$ to grow
 with the sample size $n$. When $p=p_n \to \infty$ and $q=q_n \to \infty$, it suffices to assume that  $\log(p+q)/\sqrt n \to 0$.  More generally, define $\beta_{n}^{2}=n^{-1/2}\cdot\log\max\{n,p,q\}$,
let for some $\epsilon\in(0,2)$
\begin{equation}
\ell_{n}=\max\left\{ \left(\log\max(n,p,q)\right)^{1+\epsilon},n\exp(-\beta_{n}^{-2+\epsilon})\right\} \label{elln_rate}
\end{equation}
and assume \begin{equation}
\dfrac{\log\max(n,p,q)}{\ell_{n}}\rightarrow0,\ \beta_{n}^{2}\log\dfrac{n}{\ell_{n}}\rightarrow0,\ \limsup_{n\rightarrow\infty}\dfrac{\ell_{n}}{n}<1.\label{asym_rates}
\end{equation}

For the estimation procedure described in Section~\ref{sec:estimation}, we then have
the following result on the lower bound of the confidence interval.

\begin{thm}[Tightness of the lower bound] \label{thm: lower} Under
	conditions \eqref{C2}, \eqref{C1} and \eqref{asym_rates}, for any
	sequence $t_{n}\rightarrow\infty$, $\Psi_{n}(P)<\mathrm{LB}_{n}+t_{n}n^{-1/4}\beta_{n}$
	with probability approaching 1. \end{thm}
	
Theorem \ref{thm: lower} establishes the validity and tightness of the lower bound of the confidence interval for $\tau_{\max}$. 
This  result immediately implies the asymptotic validity of our testing
procedure for $H_{0}:\ \tau_{\max}=0$ versus $H_{a}:\ \tau_{\max}>0$.
To establish the upper bound, we further assume the following margin
condition: for some sequence $t_{n}\rightarrow\infty$, there exists
a sequence of nonempty subsets $\calD_{n}^{\star}\subseteq\calD_{n}$
such that, for all $n$, 
\begin{equation}
\sup_{d_{1},d_{2}\in\calD_{n}^{\star}}\left\{ \Psi^{d_{1}}(P)-\Psi^{d_{2}}(P)\right\} =o(n^{-1/2}),\ \inf_{d\in\calD_{n}^{\star}}\Psi^{d}(P)-\sup_{d\in\calD_{n}\backslash\calD_{n}^{\star}}\Psi^{d}(P)\geq t_{n}n^{-1/2}\beta_{n}.\label{MC}
\end{equation}

\begin{thm}[Validity of the upper bound] \label{thm: upper} Under
	the same conditions as in Theorem \ref{thm: lower}, if we further
	assume \eqref{MC} or $\Psi_{n}(P)=0$ for all $n$, then $\mathrm{LB}_{n}\leq\Psi_{n}(P)\leq\mathrm{UB}_{n}$
	with probability approaching $1-\alpha$.
\end{thm}
These theorems, as well as their
 technical assumptions, are  generalizations of
 Theorems 2 and 3 of \citet{luedtke2018parametric} and specialize to their 
results when $s_{x}=1$ and $q=s_{y}=1$ in connection with the univariate maximal correlation setting of \cite{mckeague2015adaptive}.   The extension to the general multivariate analysis of variance setting (i.e., the maximal Pillai trace) is highly non-trivial because of extra challenges that arises when analyzing the canonical gradient  (given by \eqref{can1} and \eqref{CanonicalGradient}), and specifically in bounding its second-order remainder term (see the Supplementary Materials).   

\section{Greedy search for maximal Pillai trace \label{sec:algorithm} }

\subsection{Algorithm}

When computing the stabilized one-step estimator, the computationally most costly part is the optimization in \eqref{update_dnj}.  
To obtain $d_{nj}$, we  need to search over subsets $\calK$ of size $s_{x}$ and, similarly, over subsets $\calJ$  of size $s_{y}$.
This means a search over $\binom{p}{s_{x}} \binom{q}{s_{y}}$ possible combinations, which is computationally too expensive when $p$ and $q$ are large. 
In some applications, there may be a neighborhood structure that can be exploited to reduce  computational expense. For example, restricting to neighborhoods of the form $\calK=\{1,\dots,s_{x}\},\{2,\dots,s_{x}+1\},\dots$,
gives  $p-s_{x}+1$ possible subsets in total. Then in total, we only have
to compute the Pillai trace of $(p-s_{x}+1)(q-s_{y}+1)$ combinations.

Nevertheless, in general there is a need to speed up the first step of the computation of the stabilized one-step estimator given in the previous section.  To that end, we introduce the scalable greedy search in Algorithm \ref{alg:stepwise} to (approximately) maximize the Pillai trace  $\Vert\mbC_{\mbX_{\calK}\mbY_{\calJ}}\Vert_{F}^2$ 
over $\vert\calK\vert=s_x$ and $\vert\calJ\vert=s_y$.

This  algorithm is  much more efficient than a full combinatorial search.
For all $j\not\in\calJ$ and $k\not\in\calK$, we consider the increments
in the Pillai trace $\Vert\mbC_{\mbY_{\calJ}\mbX_{\calK}}\Vert_{F}^{2}$
by replacing $\calJ$ with $\calJ\cup\{j\}$ and replacing $\calK$
with $\calK\cup\{k\}$. Let $E_{j\mid\calJ}=Y_{j}-\E(Y_{j})-\bolSigma_{Y_{j}\mbY_{\calJ}}\bolSigma_{\mbY_{\calJ}}^{-1}\{\mbY_{\calJ}-\E(\mbY_{\calJ})\}$
be the residual of $Y_{j}$ regressed on $\mbY_{\calJ}$, and similarly,
$R_{k\mid\calK}$ be the residual of $X_{k}$ regressed on $\mbX_{\calK}$.
The sample versions of $E_{j\mid\calJ}$ and $R_{k\mid\calK}$ are
obtained using ordinary least squares and then plugged in the calculation
of $\mbC_{E_{j\mid\calJ}\mbX_{\calK}}$ and $\mbC_{R_{k\mid\calK}\mbY_{\calJ}}$.
This leads to the following result.

\begin{lem} \label{lem: stepwise} Assume that $\mbS_{\mbY_{\calJ}}>0$, $\mbS_{\mbX_{\calK}}>0$ and $n>\max(s_{x},s_{y})+1$.  Then 
	\begin{eqnarray}
	\Vert\mbC_{\mbY_{\calJ\cup\{j\}}\mbX_{\calK}}\Vert_{F}^{2} & = & \Vert\mbC_{\mbY_{\calJ}\mbX_{\calK}}\Vert_{F}^{2}+\Vert\mbC_{E_{j\mid\calJ}\mbX_{\calK}}\Vert_{F}^{2},\label{stepwiseY}\\
	\Vert\mbC_{\mbY_{\calJ}\mbX_{\calK\cup\{k\}}}\Vert_{F}^{2} & = & \Vert\mbC_{\mbY_{\calJ}\mbX_{\calK}}\Vert_{F}^{2}+\Vert\mbC_{\mbY_{\calJ}R_{k\mid\calK}}\Vert_{F}^{2},\label{stepwiseX}\\
	\Vert\mbC_{\mbY_{\calJ\cup\{j\}}\mbX_{\calK\cup\{k\}}}\Vert_{F}^{2} & = & \Vert\mbC_{\mbY_{\calJ}\mbX_{\calK}}\Vert_{F}^{2}+\Vert\mbC_{E_{j\mid\calJ}\mbX_{\calK}}\Vert_{F}^{2}+\Vert\mbC_{\mbY_{\calJ}R_{k\mid\calK}}\Vert_{F}^{2}+\Vert\mbC_{E_{j\mid\calJ}R_{k\mid\calK}}\Vert_{F}^{2}.\label{stepwiseXY}
	\end{eqnarray}
\end{lem}

This result gives the  increment in the Pillai
trace when including an additional variable in either $\mbX$ or
$\mbY$, or both, and allows us to implement the  greedy search 
via forward stepwise selection in Algorithm \ref{alg:stepwise}.
Another implication of Lemma \ref{lem: stepwise} is that, as we mentioned earlier in equation \eqref{eq:cca_screening}, $\max_{\vert\calK\vert=s_{x},\vert\calJ\vert=s_{y}}\Vert\mbC_{\mbY_{\calJ}\mbX_{\calK}}\Vert_{F}^{2}=\max_{\vert\calK\vert\leq s_{x},\ \vert\calJ\vert\leq s_{y}}\Vert\mbC_{\mbY_{\calJ}\mbX_{\calK}}\Vert_{F}^{2}.$
This means that even if one uses the full greedy search over $\{\vert\calK\vert\leq s_{x},\ \vert\calJ\vert\leq s_{y}\}$,
our result narrows down the search to $\{\vert\calK\vert=s_{x},\vert\calJ\vert=s_{y}\}$.

An alternative version of Algorithm \ref{alg:stepwise} involves maximizing over increments of the root-Pillai trace, which by \eqref{stepwiseY} in Lemma \ref{lem: stepwise}  can be expressed in terms of increments of the Pillai trace as
$$\Vert\mbC_{\mbY_{\calJ\cup\{j\}}\mbX_{\calK}}\Vert_{F} -\Vert\mbC_{\mbY_{\calJ}\mbX_{\calK}}\Vert_{F} = \Vert\mbC_{E_{j\mid\calJ}\mbX_{\calK}}\Vert_{F}^{2}\Big/\left(\sqrt{\Vert\mbC_{\mbY_{\calJ}\mbX_{\calK}}\Vert_{F}^{2}+\Vert\mbC_{E_{j\mid\calJ}\mbX_{\calK}}\Vert_{F}^{2}} +\Vert\mbC_{\mbY_{\calJ}\mbX_{\calK}}\Vert_{F}\right).$$ 
The above expression is an increasing function of $\Vert\mbC_{E_{j\mid\calJ}\mbX_{\calK}}\Vert_{F}^{2}$, so the same index $j$ must maximize both of these increments and the alternative version of 
Algorithm \ref{alg:stepwise} is therefore equivalent.

\begin{algorithm}[t!]
	\begin{enumerate}
		\item Initialize $\calJ=\{j\}$ and $\calK=\{k\}$, where $(j,k)$ maximizes
		$\Vert\mbC_{Y_{j}X_{k}}\Vert_{F}^{2}=\widehat{\Corr}^{2}(Y_{j},X_{k})$. 
		\item Selection over $j\not\in\calJ$ and $k\not\in\calK$. 
		\begin{enumerate}
			\item If $\vert\calJ\vert<s_{y}$ and $\vert\calK\vert<s_{x}$, find $j\not\in\calJ$
			and $k\not\in\calK$ that maximizes $\delta_{j}^{\calJ,\calK}\equiv\Vert\mbC_{E_{j\mid\calJ}\mbX_{\calK}}\Vert_{F}^{2}$
			and $\gamma_{k}^{\calJ,\calK}\equiv\Vert\mbC_{\mbY_{\calJ}R_{k\mid\calK}}\Vert_{F}^{2}$,
			respectively. Then update $\calJ\rightarrow\calJ\cup j$ if $\max_{j\not\in\calJ}\delta_{j}^{\calJ,\calK}>\max_{k\not\in\calK}\gamma_{k}^{\calJ,\calK}$,
			otherwise update $\calK\rightarrow\calK\cup k$. 
			\item If $\vert\calJ\vert<s_{y}$ and $\vert\calK\vert=s_{x}$, update $\calJ\rightarrow\calJ\cup j$
			where $j\not\in\calJ$ maximizes $\delta_{j}^{\calJ,\calK}$. 
			\item If $\vert\calJ\vert=s_{y}$ and $\vert\calK\vert<s_{x}$, update $\calK\rightarrow\calK\cup k$
			where $k\not\in\calK$ maximizes $\gamma_{k}^{\calJ,\calK}$. 
		\end{enumerate}
		\item Update the Pillai trace based on the increment given in Lemma~\ref{lem: stepwise}. 
		\item Repeat Steps 2 and 3 until $\vert\calJ\vert=s_{y}$ and $\vert\calK\vert=s_{x}$.
		\item Output: $\widehat{\calJ}$, $\widehat{\calK}$ and $\Vert\mbC_{\mbY_{\widehat{\calJ}}\mbX_{\widehat{\calK}}}\Vert_{F}^{2}$ or $\Vert\mbC_{\mbY_{\widehat{\calJ}}\mbX_{\widehat{\calK}}}\Vert_{F}$. 
	\end{enumerate}
	\caption{\label{alg:stepwise} Greedy search}
\end{algorithm}

This  greedy search algorithm is related to one proposed by \citet{wiesel2008greedy}, which was designed for sparse maximization of the sample version of the leading canonical correlation coefficient $\rho$. However, we have two important advantages. First,
due to Lemma~\ref{lem: stepwise}, we are able to maximize and update  {\it exact} increments in the Pillai trace, namely $\Vert\mbC_{E_{j\mid\calJ}\mbX_{\calK}}\Vert_{F}^{2}$
and $\Vert\mbC_{\mbY_{\calJ}R_{k\mid\calK}}\Vert_{F}^{2}$, whereas in \citet{wiesel2008greedy}
the exact increments  in $\rho$ are not available  and the maximization
is carried out on lower bounds of the increments. Second, to obtain the maximal
canonical correlation, the CCA directions $\bolalpha$ and
$\bolbeta$  also need to be updated at each step of including an additional
variable, while the update for the Pillai trace is automatically
obtained by the equations in Lemma~\ref{lem: stepwise} using linear regression residuals. Therefore, our approach is both more accurate and computationally more efficient than    \citet{wiesel2008greedy}'s  greedy search algorithm.

\begin{figure}[t!]
	\centering
\includegraphics[scale=0.66]{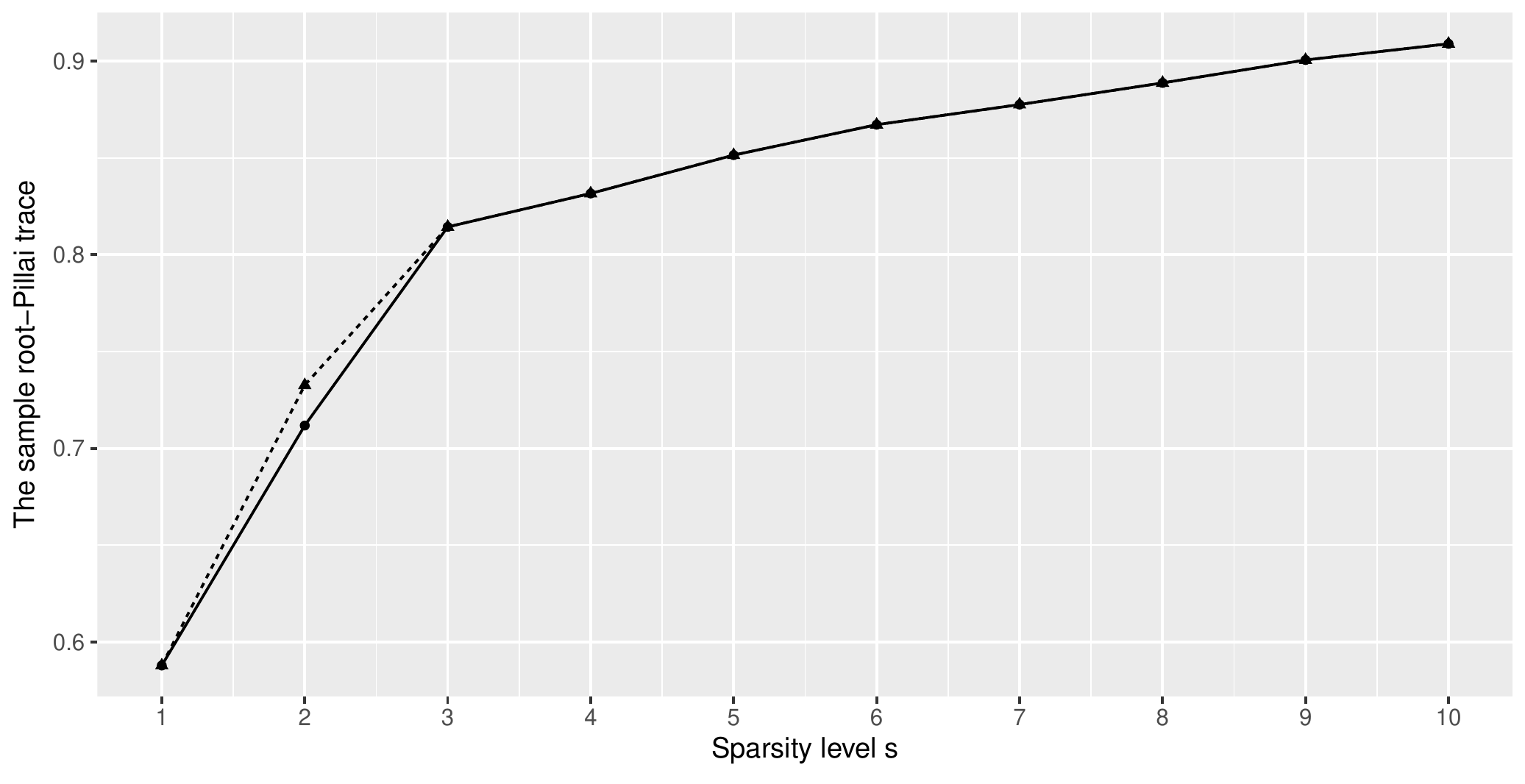}

\includegraphics[scale=0.66]{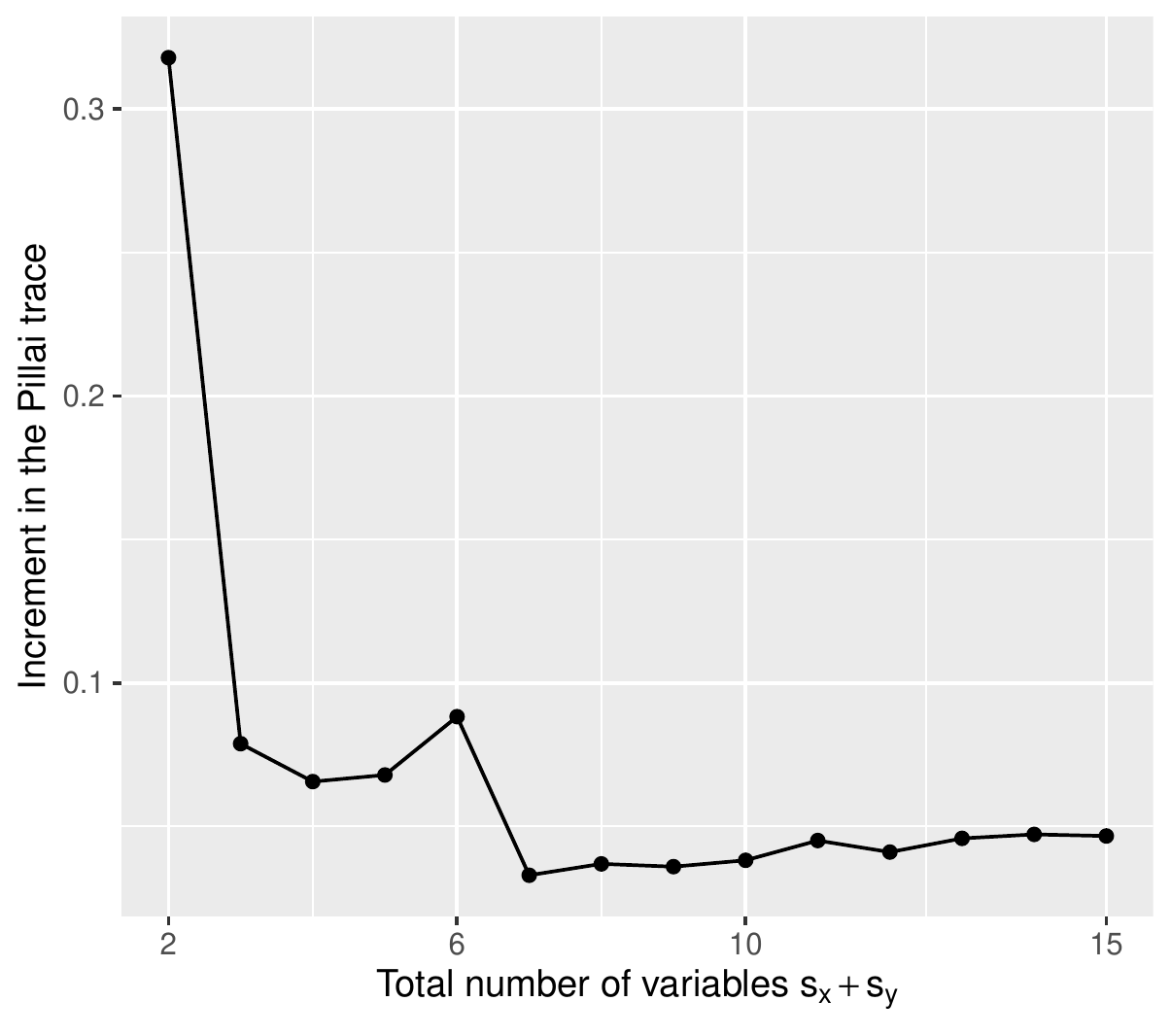}  \includegraphics[scale=0.66]{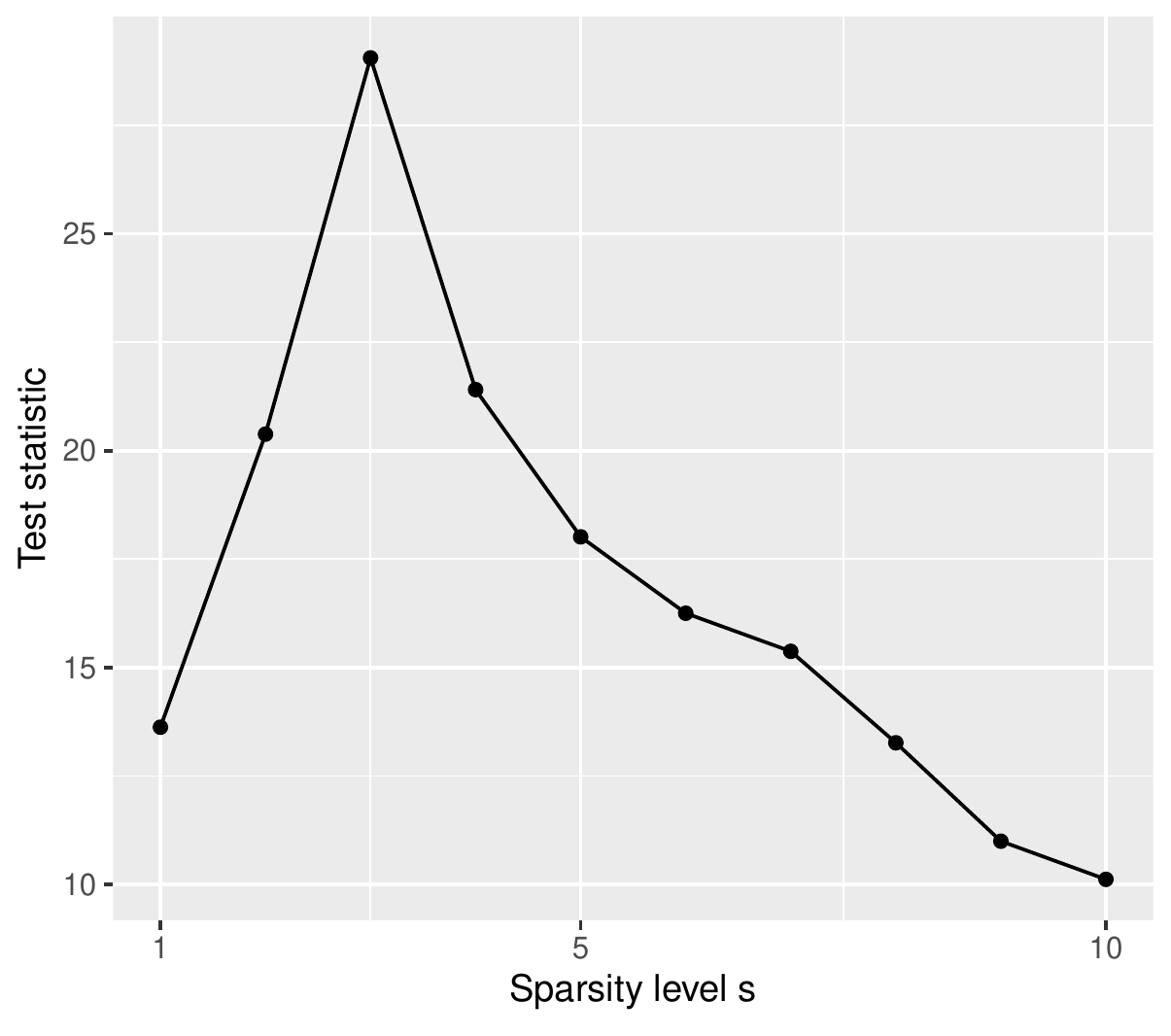}

\caption{\label{fig:forwardstep} Results based on single samples  generated under Model~A1 with $n=500$, $s_x^{\star} =s_y^{\star}=3$, $\tau_{\max}=0.8$. Top panel:  values of $\widehat{\tau}_{\rm samp}$  from a full search (dotted line) and a greedy search (solid line) as the sparsity level $s=s_x=s_y$ varies from $1$--$10$, for $p=q=10$.  Bottom left: scree plot of the increment in the sample Pillai trace, $\Vert\mbC_{E_{j\mid\calJ}\mbX_{\calK}}\Vert_{F}^{2}$ or $\Vert\mbC_{\mbY_{\calJ}R_{k\mid\calK}}\Vert_{F}^{2}$, when adding one variable at a time in the greedy search, for $p=q=5000$. Bottom right:  corresponding values of the studentized $ \widehat\tau_{\max}$ (test statistic for $\tau_{\max} =0$) as a function of $s$. }
\end{figure}

The top panel of Figure \ref{fig:forwardstep} gives the results from a toy example
  showing that the proposed greedy algorithm for finding the maximal sample root-Pillai trace under varying sparsity constraints provides almost perfect agreement with the full search. The data were generated from Model A1 used in the simulation study (in Section \ref{sec:sim}), with  $p=q=10$, the true numbers of active variables $s_x^{\star} =s_y^{\star}=3$, $\tau_{\max}=0.8$, the true number of non-zero canonical correlations $K=1$,  and $n=500$. The result of the  greedy search agrees with the full search at all sparsity levels except at $s=s_x=s_y =2$.

Algorithm~\ref{alg:stepwise} can naturally be modified so as not to require pre-specified sparsity levels. In step 2, either $j$ or $k$ is added, whichever gives the larger increment in the Pillai trace. The algorithm can then be terminated in Step 4 when the increment is smaller than some given tolerance, say $0.05$ or $0.01$.  The bottom-left panel of Figure~\ref{fig:forwardstep} shows the  successive  increments in the sample Pillai trace when adding one variable at a time (either $X_k$ or $Y_j$), using the same simulation model as the first panel except with $p=q=5000$. This plot is analogous to the scree plot used in principal component analysis and factor analysis, providing intuition and graphical diagnostics for how sparse the true model might be. It is clear from the plot that $s_x+s_y=6$ gives the most reasonable terminating point; also, at that point we have $s_x=s_y=3$ (not shown in the plot), agreeing with the true numbers of active variables.   
The bottom-right panel of Figure~\ref{fig:forwardstep} shows the corresponding values of the studentized $ \widehat\tau_{\max}$  used as the test statistic for the proposed test. 
As the sparsity level $s$ increases,  the test statistic monotonically increases until $s$ reaches the true value $s^\star_x=s^\star_y=3$, but the accuracy of the sample coherence matrix $\mbC_{\mbX_{\calK}\mbY_{\calJ}}$ used in the test statistic decreases as its dimension ($s_x\times s_y$) grows, causing a decrease in power at larger values of $s$.
In our experience, the proposed test performs well for relatively small sparsity levels, as in this simulation example, but degrades when $s^\star_x$ and $s^\star_y$ become large (say $s^\star_x=s^\star_y=20$).


\begin{figure}[t!]
	\begin{centering}
		\includegraphics[scale=0.4]{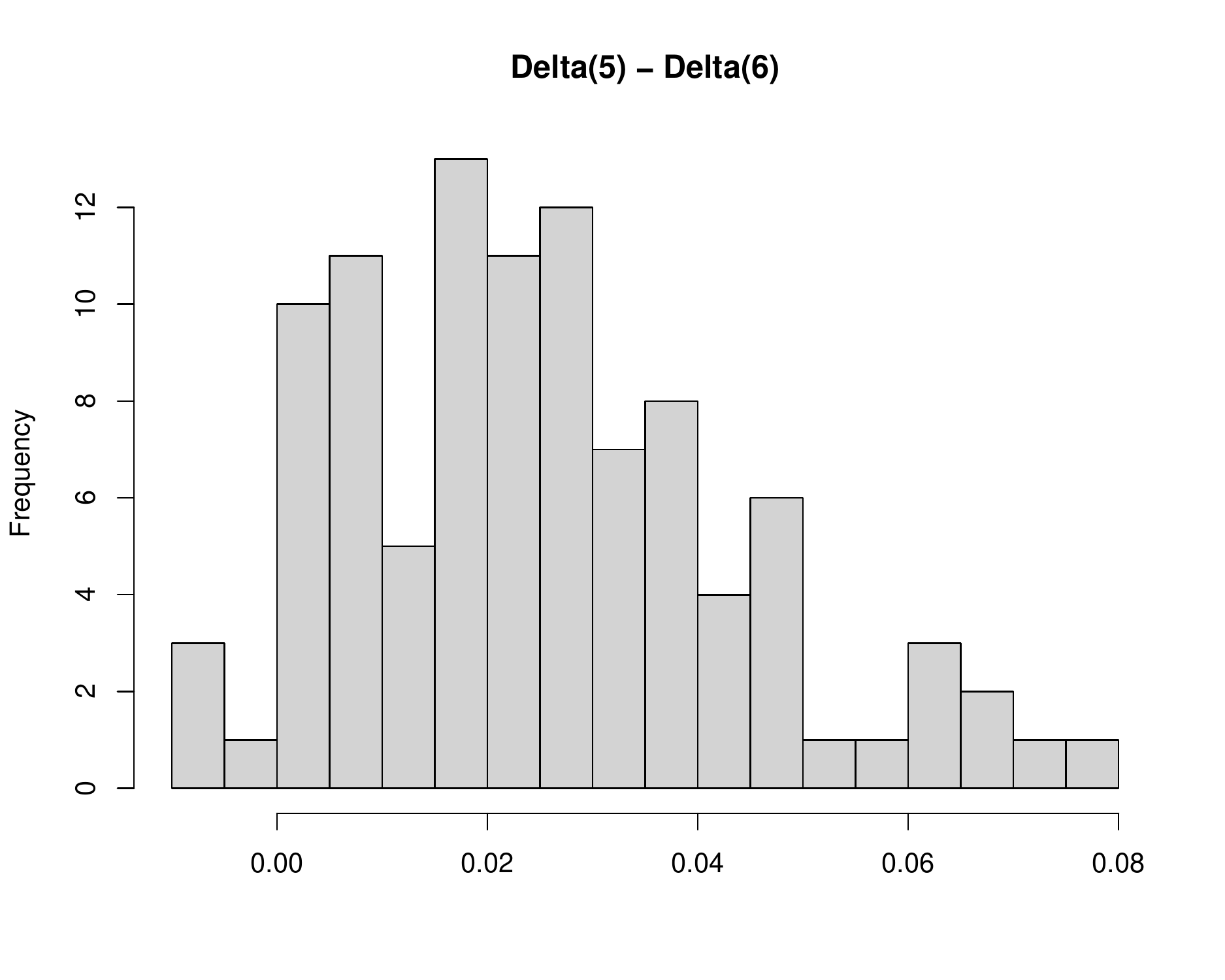}
		\includegraphics[scale=0.4]{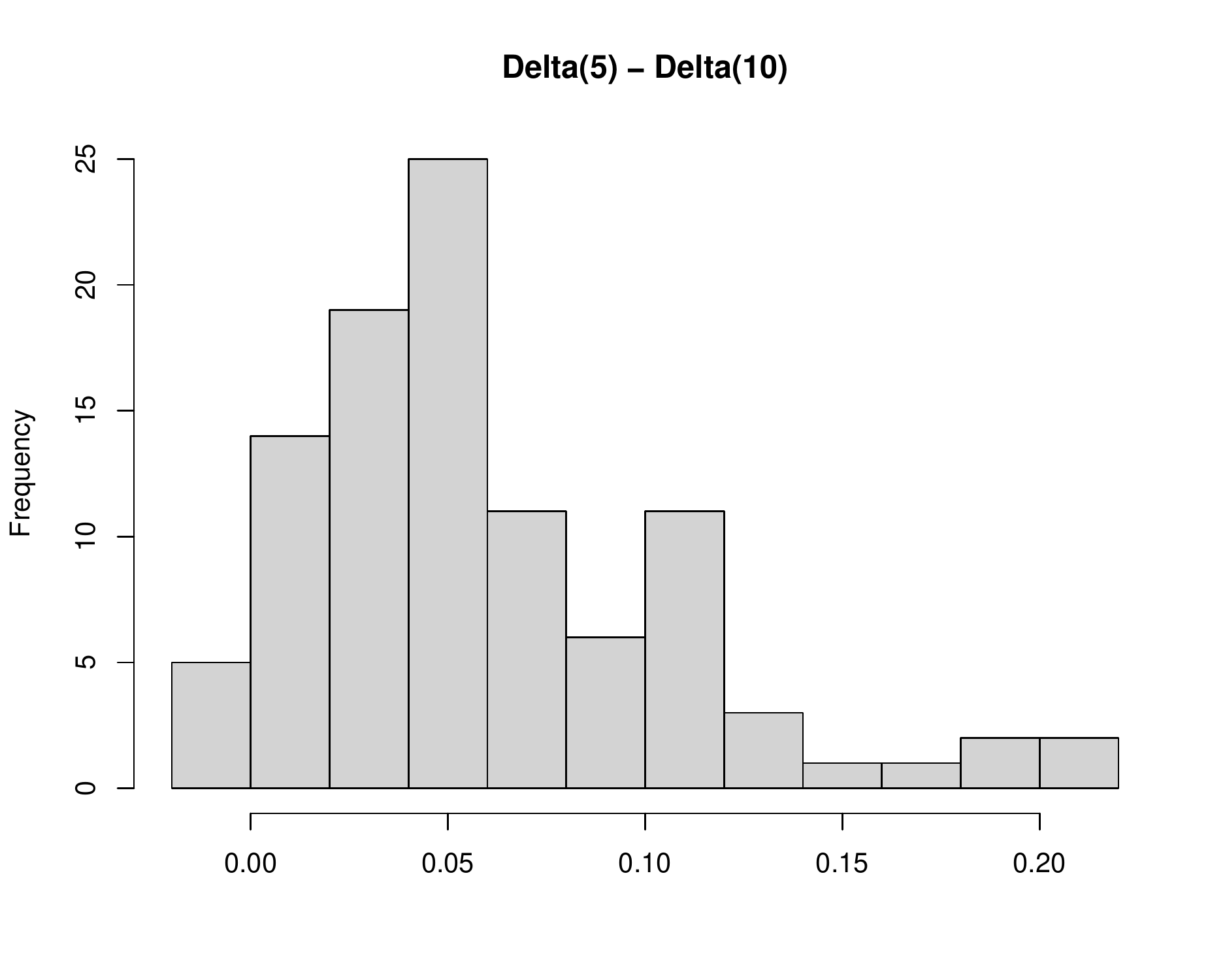}
		\par\end{centering}
	\caption{Histograms used for checking  submodularity of the root-Pillai trace as a  function  of $S=(\calJ,\calK)$ for the GBM data set.  The plotted values of $\Delta(e_i\mid S_1)-\Delta(e_i\mid S_2)$ should be non-negative if the function is submodular.  The histograms are based on 100 randomly sampled elements $e_i\not\in S_2$. The set $S_2$ is randomly chosen with $\vert\calK\vert=\vert\calJ\vert=6$ on the left, and with $\vert\calK\vert=\vert\calJ\vert=10$ on the right. For $S_1\subset S_2$ we take the first $5$ elements of $S_2$. \label{fig:submodular}}
\end{figure}

\subsection{Submodularity} To  gain some further insight into the  performance  the proposed greedy search algorithm, we  show in a simulation example that the root-Pillai trace  comes close to satisfying the submodular property (which, if true, would give  a  guarantee of finding  the  maximum to within a factor of $1/e$). Maximization of the 
root-Pillai trace over $\calK$ and $\calJ$ is a discrete combinatorial optimization problem for a set function $f\colon \ 2^{V}\mapsto \mbbR$, where the finite set $V=\{1,\dots,p\}\times \{1,\dots,q\}$ and the utility function is $f(S)=\Vert\mbC_{\mbX_{\calK}\mbY_{\calJ}}\Vert_F$ for $S=(\calJ,\calK)\subseteq V$. 
Submodularity in set functions is the discrete analogy of convexity in continuous functions. Fast greedy algorithms with theoretical guarantees have been developed for submodular function maximization \citep[see][for example]{submodular1, krause14survey, submodular2}, when the utility function is also monotonic. Specifically, the set function $f$ is monotonic if for any two sets $S_1\subseteq S_2\subseteq V$, we  have $f(S_1)\le f(S_2)$. Based on Lemma~\ref{lem: stepwise}, it is not difficult to see that $f(S)=\Vert\mbC_{\mbX_{\calK}\mbY_{\calJ}}\Vert_F$ is monotonic.  

We now briefly review the definition of submodularity  and then numerically demonstrate that our set function $f$ can be close to submodular, although not exactly so. 
A key concept related to submodularity is the discrete derivative. The discrete derivative of $f$ at $S$ with respect to a new element $e\in V$ is defined as $\Delta(e\mid S)=f(S\cup\{e\}) - f(S)$. Then $f$ is called submodular if, for any $S_1\subseteq S_2\subset V$ and $e\in V\setminus S_2$, we have $\Delta(e\mid S_1)\geq \Delta(e\mid S_2)$. We consider a simple numerical experiment using the real data in Section~\ref{sec:real}, where $(p,q,n)=(1000,534,397)$. First, we randomly sampled $S_2\subset V$ with size $s_x=s_y=6$ (or 10) and defined the first $5$ elements in $S_2$ to be $S_1$. Then we computed the histogram of the differences $\Delta(e_i\mid S_1) - \Delta(e_i\mid S_2)$ for 100 randomly selected elements $e_i\not\in S_2$. The results  displayed in Figure~\ref{fig:submodular} indicate a close approximation to submodularity since there are very few negative values in each histogram.  In contrast, the Pillai trace is readily seen to violate the submodular property: the scree plot in  Figure  \ref{fig:forwardstep} is not  monotonically decreasing. 

\section{Simulation study\label{sec:sim}}


 The sample size is fixed at $n=500$, while we vary the dimensions of $\mbX$ and $\mbY$ from $p=q=10$ to $p=q=5000$. We generated i.i.d.~samples $(\mbX_{i}^{T},\mbY_{i}^{T})^{T}\in\mbbR^{p+q}$, $i=1,\dots,n$, from a joint normal distribution with mean
zero and covariance specified by
\begin{equation}\label{sim:cov}
	(\bolSigma_{\mbX})_{jl}=(\bolSigma_{\mbY})_{jl}=\begin{cases}
		0.5^{\vert j-l\vert}, & j,l\leq100,\\
		I(j=l), & \mathrm{otherwise},
	\end{cases}\quad\quad \bolSigma_{\mbX\mbY}=\bolSigma_{\mbX}\left(\sum_{k=1}^{K}\rho_{k}\bolalpha_{k}\bolbeta_{k}^{T}\right)\bolSigma_{\mbY}.
\end{equation}
The above structured $\bolSigma_{\mbX\mbY}$ is commonly used in the sparse CCA literature \citep[e.g.,][]{mai2019iterative}, where $K$ is the number of non-zero CCA coefficients, $\rho_{k}>0$
is the $k$-th canonical correlation, $\bolalpha_{k}$ and $\bolbeta_{k}$ are the corresponding sparse CCA directions that satisfy all the length, orthogonality and sparsity constraints. Also, the maximal canonical correlation coefficient $\rho=\rho_1$. Note that the number $K$ is irrelevant in our estimation as we did not use that information. Under this simulation setting, the covariance matrices $\bolSigmaX$, $\bolSigmaY$ and $\bolSigmaXY$ are not sparse while the sparsity is imposed directly on each $\bolalpha_k$ and $\bolbeta_k$. The nonzero elements in $\bolalpha$ and $\bolbeta$ correspond to the active variables in $\mbX$ and $\mbY$, respectively. In our simulations, we have the symmetry in $\mbX$ and $\mbY$ and thereby set $\bolalpha_k=\bolbeta_k$, which implies $s^\star_x=s^\star_y$.

We consider three scenarios of the form \eqref{sim:cov}.
The first scenario is a model satisfying null hypothesis (Model N), where $\bolSigma_{\mbX\mbY}=0$,  $K=0$,  $s^{\star}_{x}=s^{\star}_{y}=0$, and  we vary the prescribed sparsity levels $s_x=s_y=s\in\{1,2,3,4\}$. 
The next two scenarios are  alternative hypothesis models (Models A1 and A2), with the true numbers of active variables  $s^{\star}_{x}=s^{\star}_{y}=3$.  Without loss of generality, the active variables are taken as the  first three components of $\mbX$ and  $\mbY$.  Model A1 is the single pair CCA model with $K=1$, so $\tau=\rho$. The SCCA direction $\bolalpha_1=\bolbeta_1$ is set as $\mbv_1/\sqrt{\mbv_1^T\bolSigma_{\mbX}\mbv_1}$ to satisfy the length constraint, where $\mbv_1 = (1,1,1,0,\dots,0)^T$. 
For Model A2, a general SCCA model, we take  the
number of components $K=3$ and set $(\rho_1,\rho_{2},\rho_{3})=(\tau,2\tau,3\tau)/\sqrt{14}$. The SCCA directions $\bolalpha_k=\bolbeta_k$, $k=1,2,3$, are set to have 1 in the $k$-th component and 0's elsewhere. 
Under the Models A1 and A2, we  vary $\tau\in\{0.1, 0.2, 0.3, 0.4\}$ to study the effect of changes in  the strength of the correlation.

\begin{table}[t]
	\centering
	\resizebox{\textwidth}{!}{
		{\renewcommand{\arraystretch}{1.5}	
			\begin{tabular}{llrrrrrrrrrrrrl}
				\hline
				\toprule
				\multirow{1}{*}{} &  & \multicolumn{4}{c}{Model N}  & \multicolumn{4}{c}{Model A1} & \multicolumn{4}{c}{Model A2}\tabularnewline
				$p$ & & {$s=1$} & $2$ & $3$ & $4$ & $\tau=0.1$ & $0.2$ & $0.3$ & $0.4$ & $\tau=0.1$ & $0.2$ & $0.3$ & $0.4$\tabularnewline
				\toprule
				\multirow{5}{*}{$10$} & OS & 0.066 & 0.056 & 0.050 & 0.064 & 0.124 & 0.546 & 0.950 & 1 & 0.098 & 0.448 & 0.894 & 0.998\tabularnewline
				& HC & 0.122 & 0.530 & 0.542 & 0.532 & 0.802 & 0.980 & 1 & 1 & 0.734 & 0.960 & 1 & 1\tabularnewline
				& MF & 0.050 & 0.050 & 0.050 & 0.050 & 0.096 & 0.368 & 0.860 & 0.996 & 0.104 & 0.388 & 0.888 & 0.998\tabularnewline
				& SF & 0.984 & 0.952 & 0.930 & 0.894 & 0.962 & 0.998 & 1 & 1 & 0.984 & 1 & 1 & 1\tabularnewline
				& BF & 0.050 & 0.004 & 0.004 & 0 & 0.020 & 0.312 & 0.936 & 1 & 0.016 & 0.254 & 0.846 & 1\tabularnewline
				\midrule
				\multirow{4}{*}{$30$} & OS & 0.058 & 0.054 & 0.074 & 0.056 & 0.068 & 0.312 & 0.830 & 0.996 & 0.064 & 0.234 & 0.720 & 0.980\tabularnewline
				& HC & 0.134 & 0.628 & -- & -- & -- & -- & -- & -- & -- & -- & -- & --\tabularnewline
				& MF & 0.034 & 0.034 & 0.034 & 0.034 & 0.040 & 0.072 & 0.216 & 0.470 & 0.042 & 0.074 & 0.216 & 0.496\tabularnewline
				& BF & 0.048 & 0.008 & 0.002 & 0 & 0.002 & 0.074 & 0.662 & 0.996 & 0.004 & 0.062 & 0.530 & 0.966\tabularnewline
				\midrule
				\multirow{2}{*}{$100$} & OS & 0.054 & 0.072 & 0.070 & 0.080 & 0.074 & 0.190 & 0.660 & 0.982 & 0.058 & 0.136 & 0.588 & 0.946\tabularnewline
				& BF & 0.046 & 0.010 & 0.002 & 0 & 0.002 & 0.018 & 0.366 & 0.962 & 0 & 0.024 & 0.304 & 0.906\tabularnewline
				\midrule
				\multirow{2}{*}{$1000$} & OS & 0.066 & 0.056 & 0.050 & 0.064  & 0.066 & 0.076 & 0.274 & 0.866 & 0.072 & 0.074 & 0.334 & 0.838\tabularnewline
				& BF & 0.046 & 0.010 & 0.002 & 0 & 0.002 & 0.002 & 0.072 & 0.664 & 0.002 & 0.002 & 0.068 & 0.646\tabularnewline
				\midrule
				\multirow{2}{*}{$5000$} & OS & 0.082 & 0.068 & 0.072 & 0.066 & 0.072 & 0.076 & 0.154 & 0.670 & 0.066 & 0.072 & 0.174 & 0.732\tabularnewline
				& BF & 0.052 & 0.002 & 0 & 0 & 0 & 0 & 0.016 & 0.330 & 0 & 0 & 0.016 & 0.428\tabularnewline
				\bottomrule 			\hline
			\end{tabular}
		}
	}	
	\protect\caption{\label{tab:sim} Simulation under the null Model N and the two alternative models A1 and A2. 
	The reported numbers are the rejected proportion based on $500$ replicated data sets for each of the simulation settings.
	}
\end{table}

\subsection{Simulation results for hypothesis testing}
 
We compared various methods (whenever they are applicable) for the $5\%$-level test of $\tau_{\max}=0$ versus $\tau_{\max}>0$: the proposed testing procedure using the stabilized one-step estimator (OS); the classical F-test for the Pillai trace without variable selection, as implemented in the {\tt manova} \textsf{R} package (MF); naive application of the  F-test on selected variables (SF), which comes without any adjustment for variable selection; the F-test on selected variables with Bonferroni correction (BF); and the Higher Criticism (HC) method \citep{donoho2004hc,donoho2015hc} based on p-values computed from the F-test for all $\binom{p}{s_{x}} \binom{q}{s_{y}}$ combinations of variables. The HC statistic was calculated following the procedure described in \citet[Sections 1.1 and  2.1]{donoho2015hc} with the critical value calculated using the Gumbel distribution. For all methods that require variable selection (SF, BF, and OS), the variables were selected using Algorithm \ref{alg:stepwise}. All of the F-tests (MF/SF/BF) considered, as well as the HC procedure, are  based on  p-values for the MANOVA F-test that targets the Pillai trace, whereas our approach  targets the root-Pillai trace.  {For the BF procedure, although we only used the F-statistic based on the $s_x+s_y$ variables selected from Algorithm~\ref{alg:stepwise}, the Bonferroni correction covers all $\binom{p}{s_{x}} \binom{q}{s_{y}}$ combinations of variables potentially involved in the F-test.}

In Table~\ref{tab:sim}, we report the proportion of rejections under each simulation setting,  based on 500 simulation replications for each case. For the HC procedure, the total number of test statistics in one replication is $\binom{p}{s_{x}} \binom{q}{s_{y}}$. Therefore, it was only included for $p=10$ and for  $p=30, s<3$ scenarios, and was shown to have unsatisfactory type I error control. The MANOVA F-test (MF) worked well for low-dimensional null and alternative models, but is not applicable for $p>30$. When $p\geq 30$, the F-test based on selected variables without any adjustment (SF) will always reject the null even under the null hypothesis (not show this in Table~\ref{tab:sim}). This is not surprising as it fails to adjust for  spurious correlations. The only two feasible methods for high-dimensional settings are seen to be  the proposed test based on the stabilized one-step estimator (OS) and the Bonferroni corrected F-test (BF). Clearly, the proposed method has much better type I error control (under Model N) and much smaller type II error (under Models A1 and A2) than BF.  {Overall, the proposed OS testing procedure has adequately controlled the type I error around the nominal level $\alpha=0.05$. Specifically, the type I error is always between $0.05$ and $0.1$ for all different $p\in\{10,30,100,1000,5000\}$ and $s\in\{1,2,3,4\}$ combinations. Thought the test procedure is asymptotically valid, the slightly anti-conservative results appear to be caused by the stabilized one-step estimation procedure at small sample size $j< n$. In contrast, HC and SF fail to control the type I error; BF is too conservative when $s>1$; MF has the perfect type I error control only for $p=10$, which verifies the superiority of the F-test over the Wald-type test in low-dimensional settings. It is also very encouraging to see that the proposed OS testing procedure is more powerful than BF and MF, even in low dimensions ($p=10, 30$), and is able to detect weak signals (i.e.~canonical correlations are no larger than $0.4$ in all models) in very high dimensions ($p=1000, 5000$).   }

\begin{figure}[ht!]
	\begin{centering}
		\includegraphics[scale=0.47]{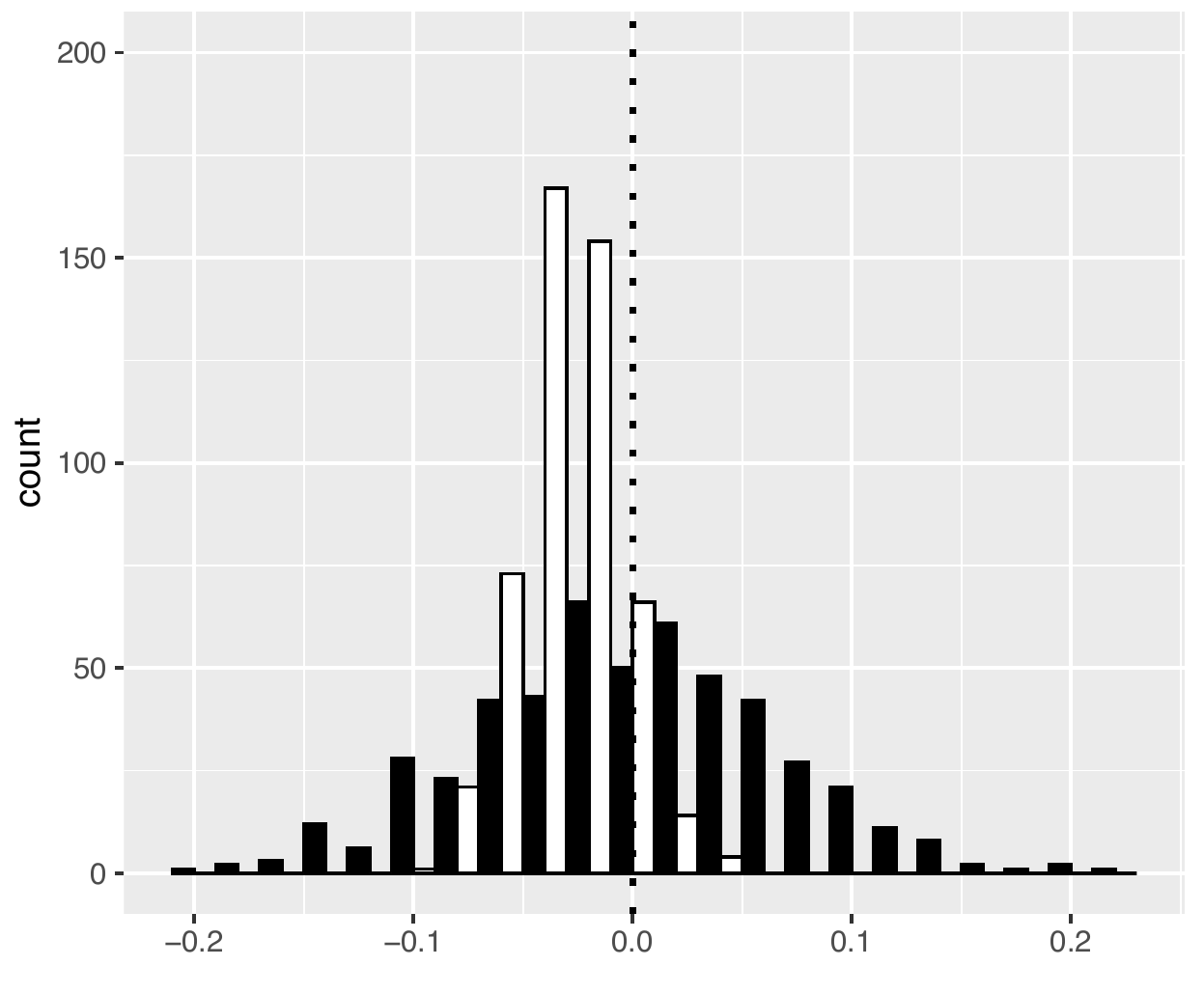}
		\includegraphics[scale=0.47]{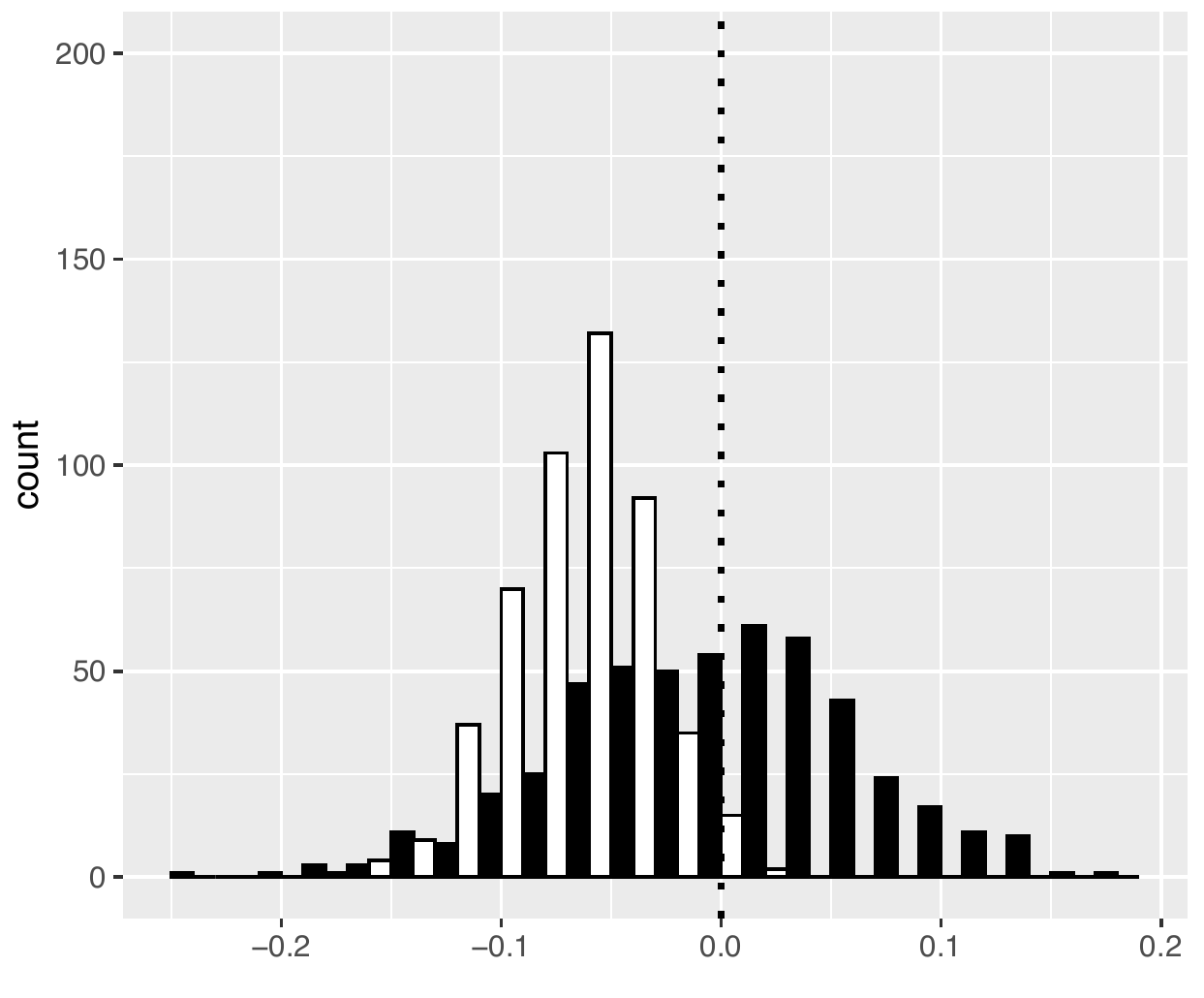}
		\includegraphics[scale=0.47]{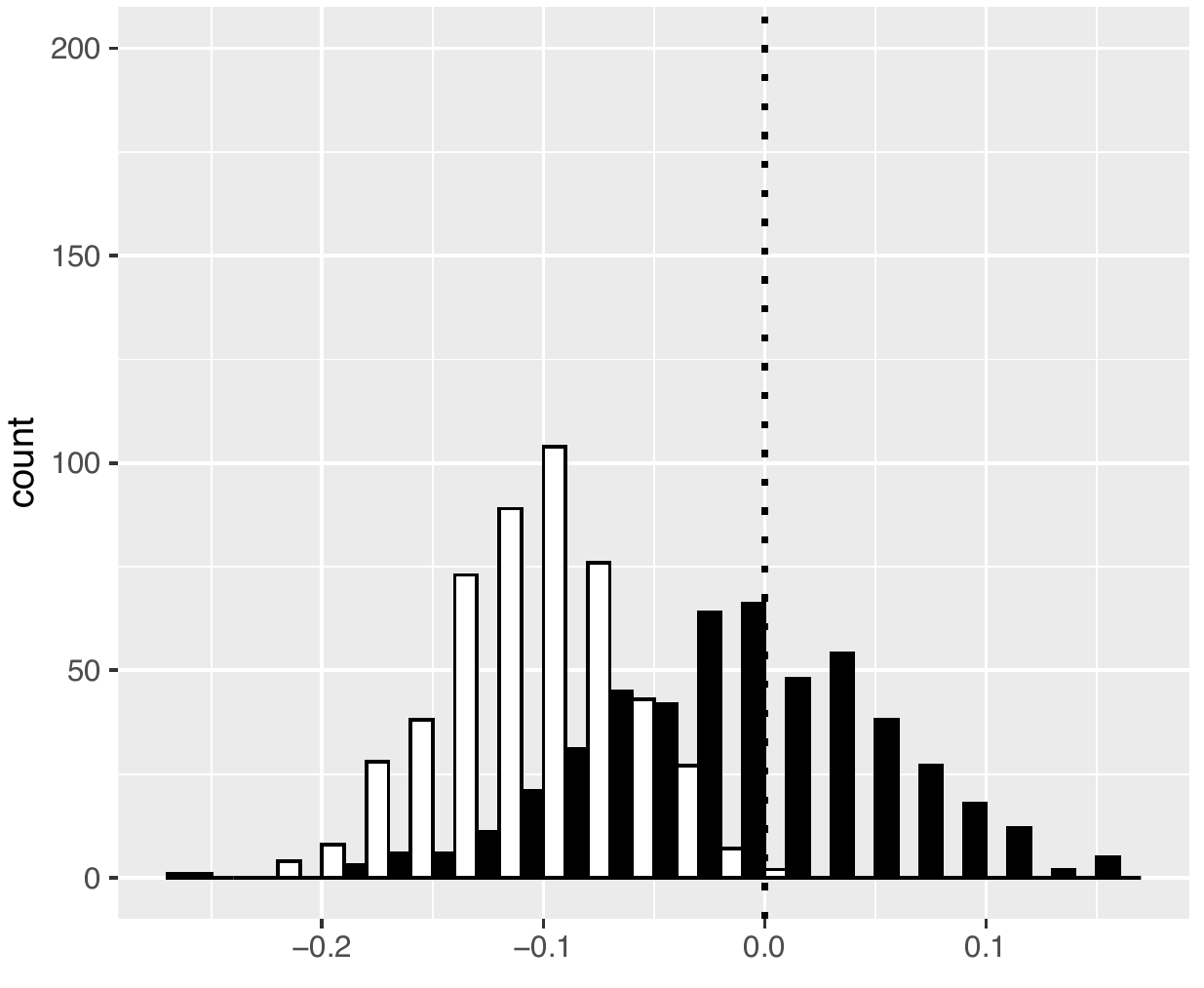}
		\includegraphics[scale=0.47]{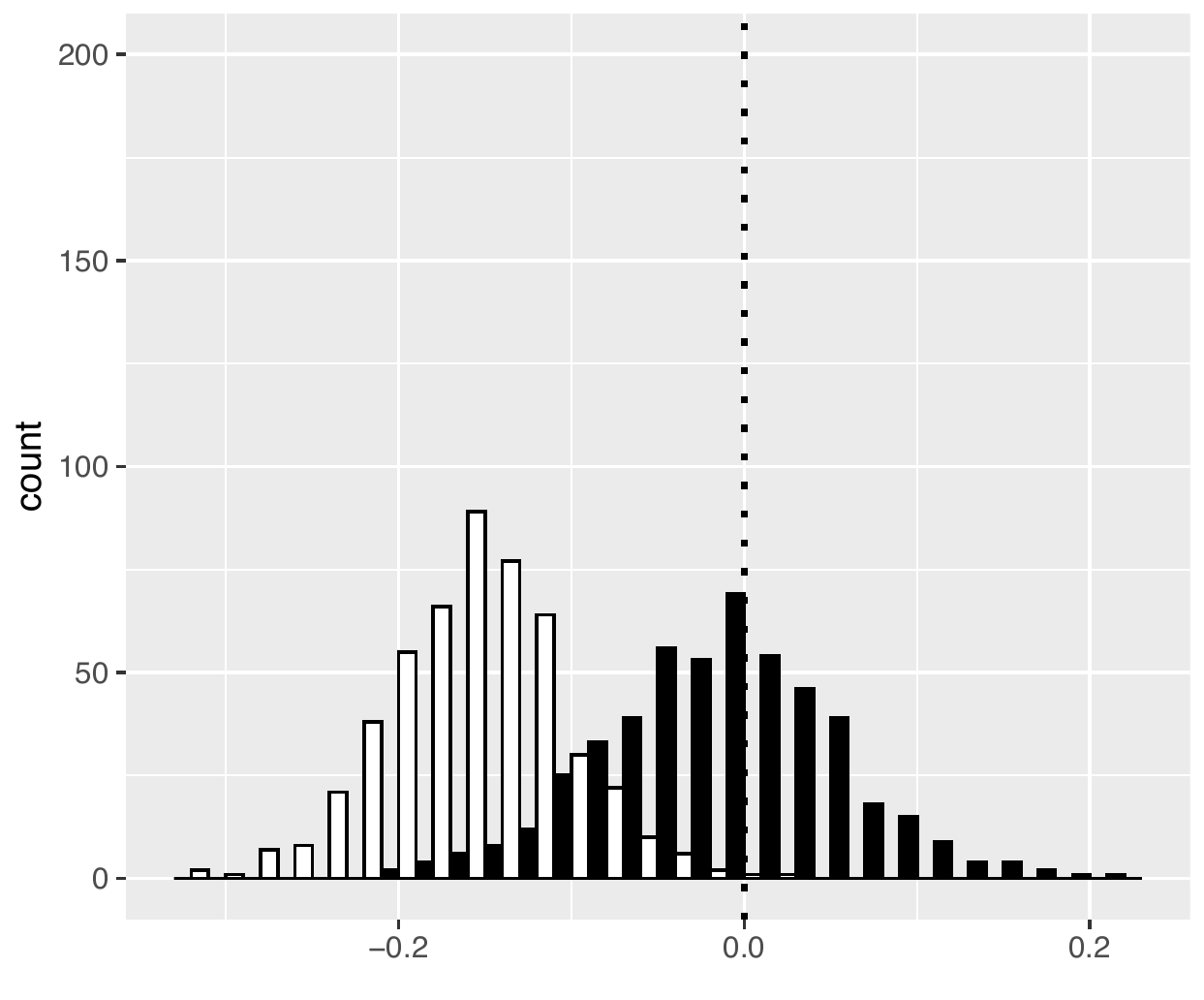}
		\par\end{centering}
	\caption{Histograms of $\widehat\tau_{\max}$  (in black) under the null  Model N based on 500 independent simulated data sets. The four panels, from top-left to bottom-right, correspond to the sparsity levels $s_x=s_y=s=1,2,3$ and $4$, respectively. The histograms in white   are of  the stabilized one-step estimator of $\tau_{\max}^2$, showing a negative bias that becomes increasingly pronounced as $s$ increases. }\label{fig:null}
\end{figure}

\begin{figure}[ht!]
	\begin{centering}
		\includegraphics[scale=0.5]{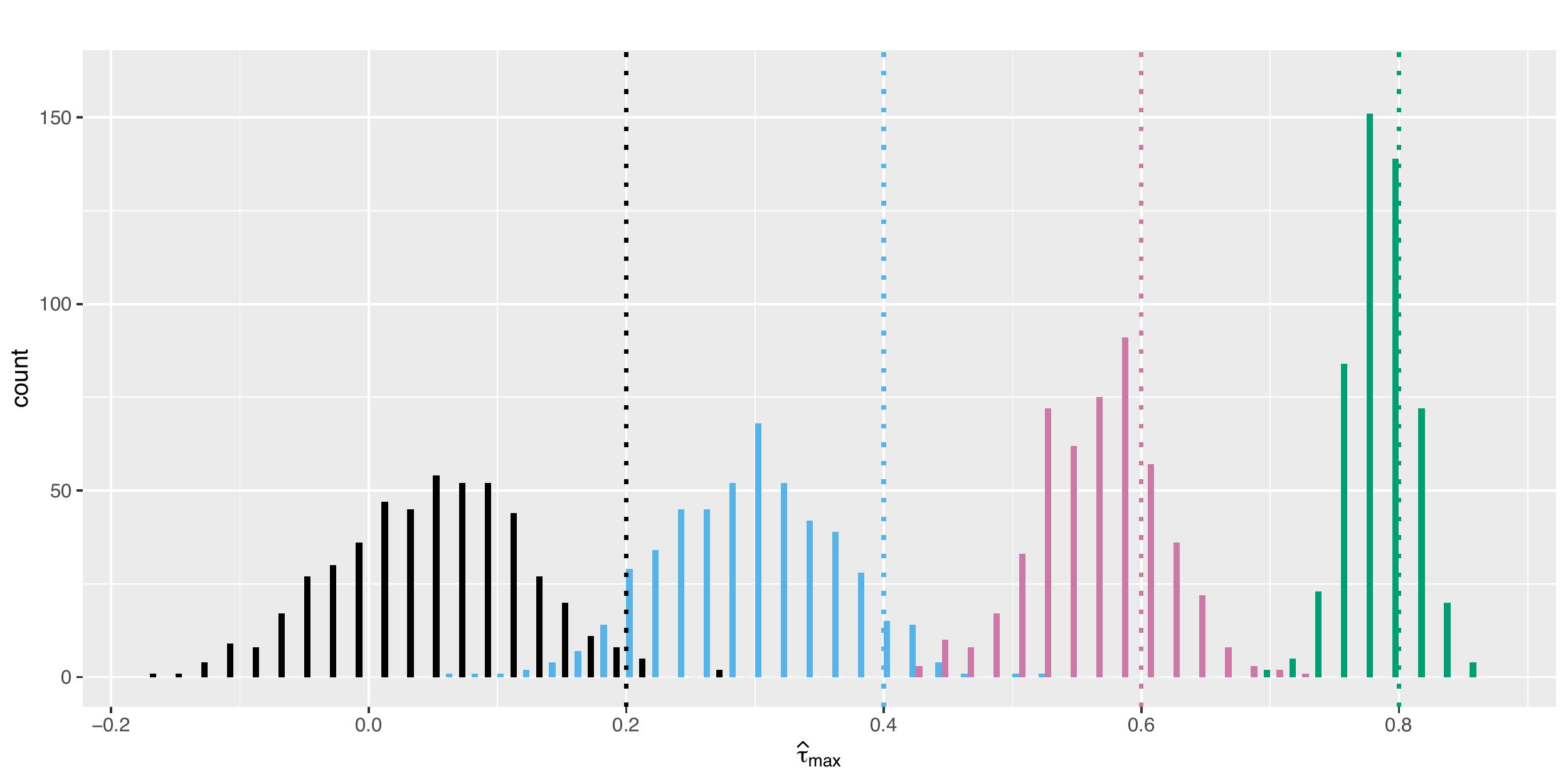}
		\includegraphics[scale=0.5]{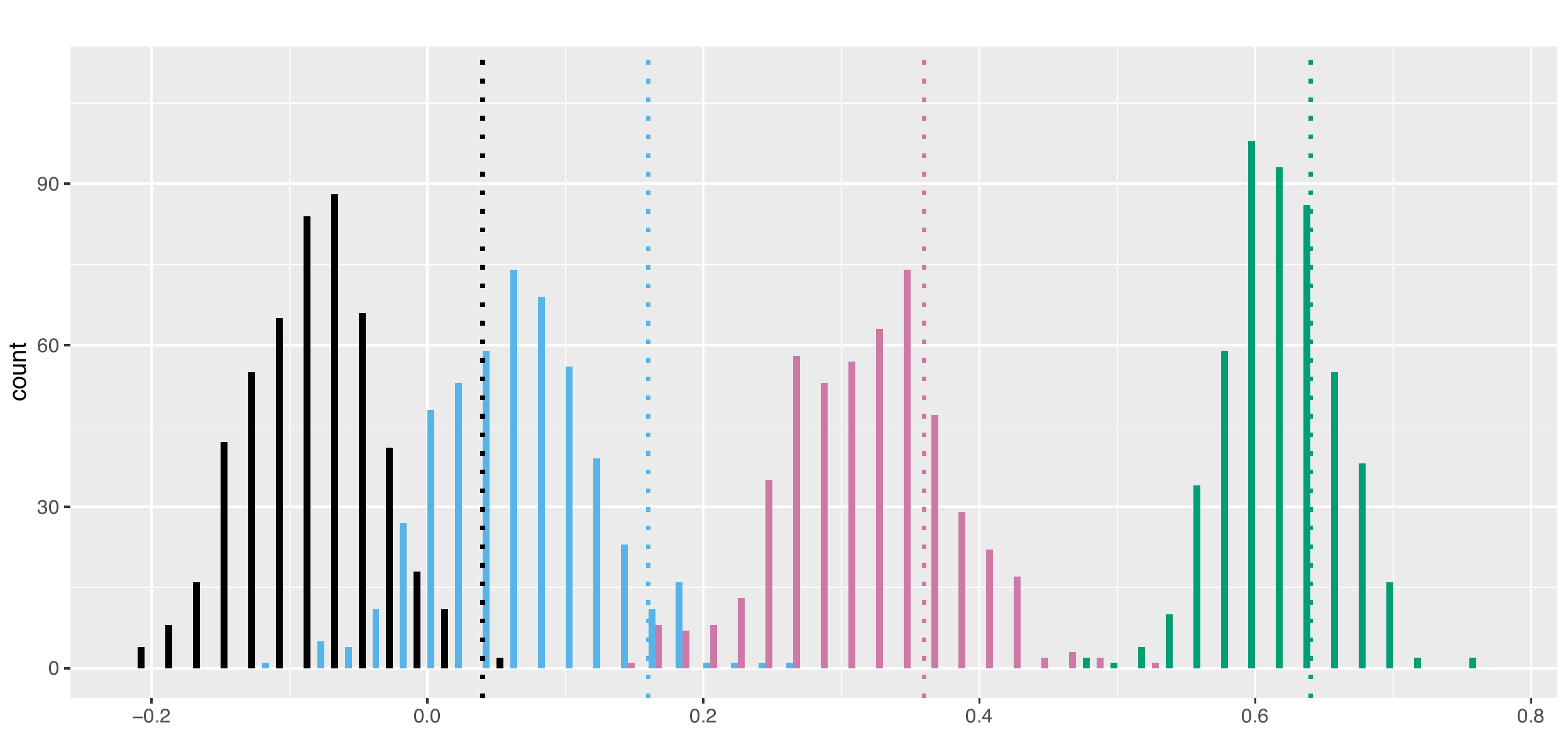}
		\par\end{centering}
	\caption{Histograms of the stabilized one-step estimators for $\tau_{\max}$ (top panel) and $\tau_{\max}^2$ (bottom panel), under Model A1.  In each plot, the four colored histograms going from left to right correspond to $\tau_{\max}=0.2$, $0.4$, $0.6$ and $0.8$, respectively. The vertical dashed lines are the true values of the targeted parameters $\tau_{\max}$ (top) and $\tau_{\max}^2$ (bottom).  }\label{fig:alt}
\end{figure}

\subsection{Simulation results for parameter estimation}\label{hyp-sim}

Although our theory and implementation are equally applicable to  stabilized one-step estimators of $\tau_{\max}$ and $\tau_{\max}^2$, the empirical results for $\tau_{\max}$ are generally better than those of $\tau_{\max}^2$. Note that the stabilized one-step estimator of $\tau^2_{\max}$ is not simply $\widehat{\tau}_{\max}^2$. 

Histograms of the estimated $\tau_{\max}$ and $\tau_{\max}^2$ from 500 independent samples (of size $n=500$) under the null Model N are presented in Figure~\ref{fig:null}, where we vary $s_x=s_y=s \in\{1,2,3,4\}$.  The stabilized one-step estimates for $\tau_{\max}$ and $\tau_{\max}^2$ are both seen to be approximately normal. For $\tau_{\max}$, the estimates are all centered around the truth, $\tau_{\max}=0$, regardless of the choice of $s$. However, for $\tau_{\max}^2$, there is a severe under-estimation phenomenon, which becomes more pronounced as  $s$ increases. We think there are two factors contributing to this phenomenon. First, the number of parameters  in $\mbC_{\mbY_{\calJ}\mbX_{\calK}}$ is $s^2$. This  requires a larger sample size $n$ for the asymptotic properties to come into effect as $s$ increases. Second, when the  Pillai trace is close to zero, its absolute value is magnified by taking the square root, making it easier to estimate. 

Under the alternative model (Model A1) with the correlation strength $\tau_{\max}$ varying from 0.2 to 0.8,  the histograms of $\widehat\tau_{\max}$ (again based on 500 independent samples of size $n=500$) are given in the top panel of Figure~\ref{fig:alt}.  Recall that the true sparsity levels are $s_x^\star=s_y^\star=3$ in this model. We also set  $s_x=s_y=s=3$.   Although there is an issue of under-estimation for both $\tau_{\max}$ and $\tau_{\max}^2$ when the signal is weak ($\tau_{\max}= 0.2$ and $0.4$),  there is a substantial improvement when the correlation is strong enough. The improvement is more pronounced in the histogram of $\widehat\tau_{\max}$ compared with that of the stabilized one-step estimator of $\tau_{\max}^2$ (bottom panel). It is worth noting that $\tau_{\max}=0.8$ is still relatively weak correlation (e.g., the estimated $\tau_{\max}$ exceeds $1.5$ in the real data example of Section \ref{sec:real}), but both estimators worked well at $\tau_{\max}=0.8$.  An explanation for the under-estimation is that the stabilizing procedure tends to attenuate the estimates to some extent, at least in the neighborhood of $\tau_{\max}=0$.  However, as seen in Figure ~\ref{fig:null}, the behavior of $\widehat\tau_{\max}$ under the null model is unaffected by such attenuation, being approximately zero-mean~normal.

\section{Analysis of glioblastoma multiforme data\label{sec:real}}

\begin{figure}[t!]
	\begin{centering}
		\includegraphics[scale=0.55]{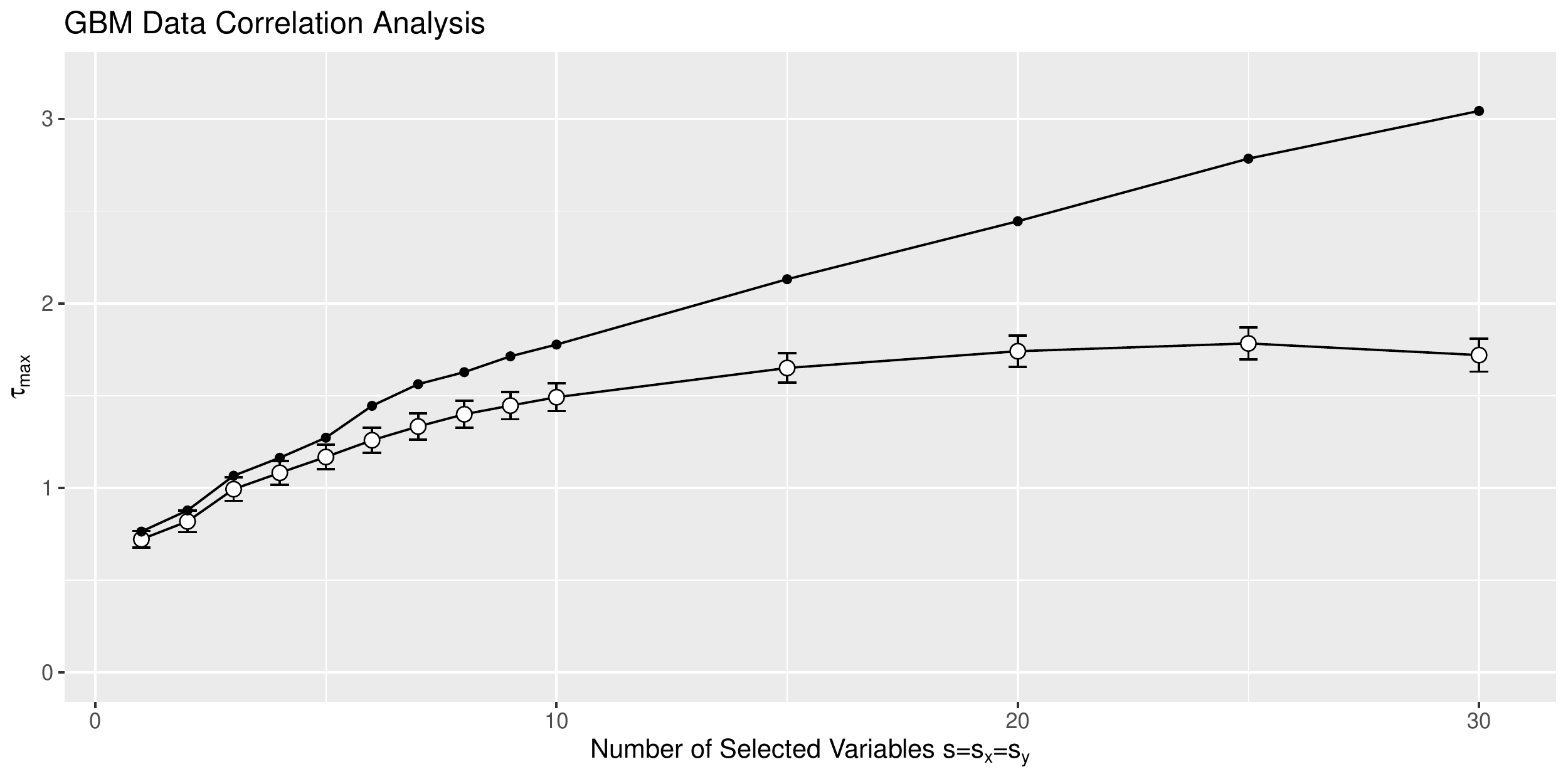}
		\par\end{centering}
	\caption{GBM data analysis. Estimates of the maximal root-Pillai trace $\tau_{\rm max}$ are plotted against $s=s_x=s_y$ varying from $1$--$30$, with the selected variables at each step found using Algorithm~\ref{alg:stepwise}. The black dots are $\widehat\tau_{\rm samp}$ (without adjustment for post-selection),  giving inflated estimates of $\tau_{\rm max}$. The white dots and associated 95\% confidence intervals are based on the stabilized one-step estimator $\widehat\tau_{\max}$, with 10 random re-orderings and averaged point estimates and averaged lower/upper CI endpoints over these re-orderings. } \label{fig:GBM}
\end{figure}

Glioblastoma multiforme (GBM) is a type of fast-growing brain tumor that is also the most common primary form of brain tumor in adults. Data were  collected by The Cancer Genome Atlas project \citep[TCGA][]{weinstein2013cancer} on $490$ patients with GBM, including  data on $q=534$ microRNA expression  and $17,472$ gene expression measurements for each patient.  It is of interest to find associations between  microRNA  and gene expression.  Following previous studies \citep{wang2015joint, molstad2019insights}, we analyze the $p=1000$ genes  with the largest median absolute deviations in gene expression, and preprocess the data by 
removing 93 subjects whose gene expression is substantially different from the majority. The resulting sample size in our data analysis is then $n=397$.

\begin{figure}[t!]
	\begin{centering}
		\includegraphics[scale=0.45]{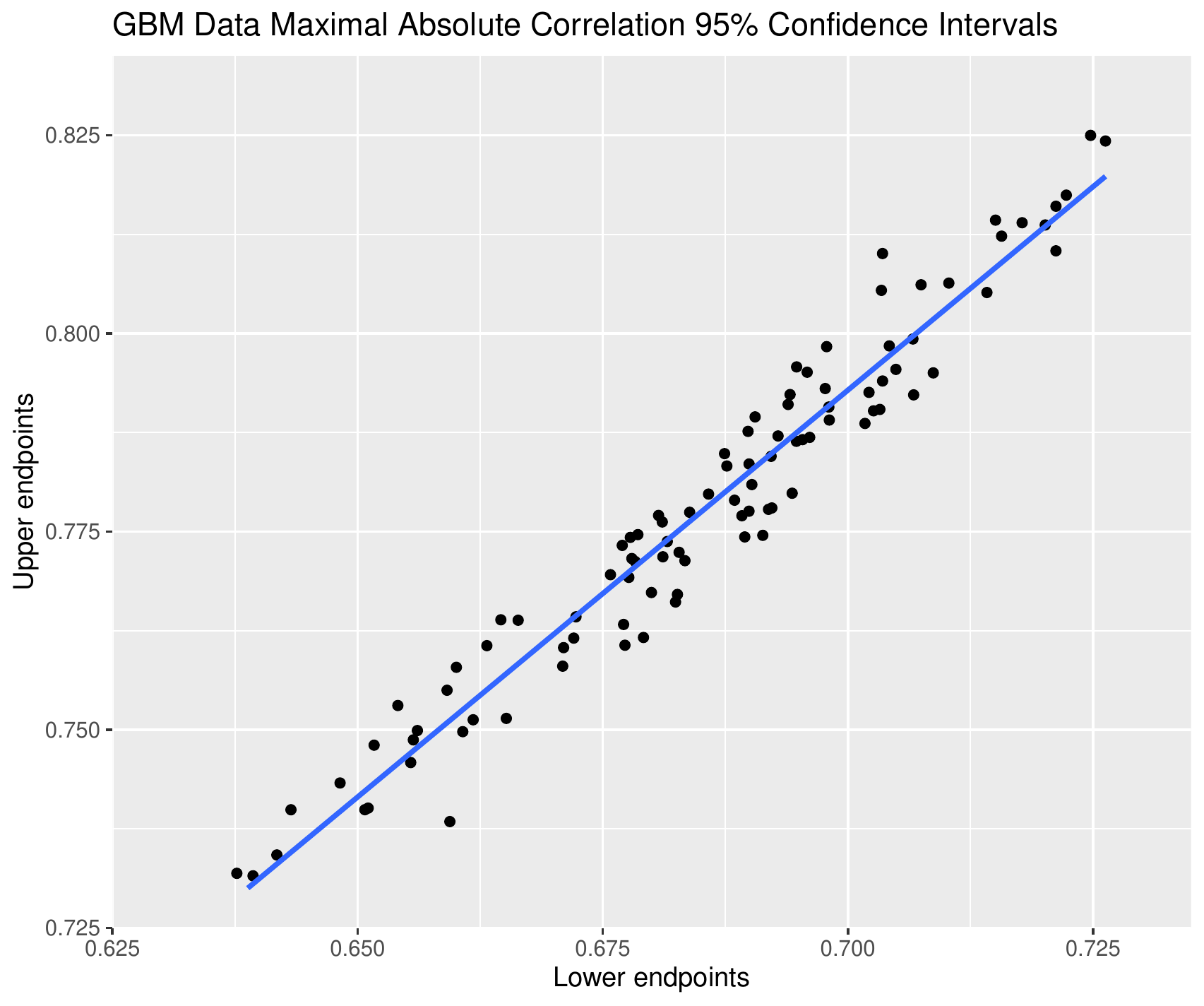}
		\par\end{centering}
	\caption{ GBM data analysis. The  lower and upper endpoints of the $95\%$ confidence interval for $\tau_{\max}$ when $s=1$, for 100 random re-orderings of the data.  }\label{fig:GBM100}
\end{figure}

We applied our Algorithm~\ref{alg:stepwise} to this data set and obtained estimates of the maximal root-Pillai trace $\tau_{\rm max}$ over a range of values of  $s=s_x=s_y$. The results are displayed in Figure~\ref{fig:GBM}. Without adjusting for the post-selection, the sample estimate $\widehat\tau_{\rm samp}$ of $\tau_{\rm max}$ increases almost linearly due to spurious correlations. On the other hand, the stabilized one-step estimator $\widehat\tau_{\max}$ gives  reasonable estimates that settle down beyond $s= 15$. The confidence intervals suggest that there is a highly significant association between  microRNA and  gene expression (with p-value less than $10^{-10}$), which is  consistent with previous studies. 
The results for the stabilized one-step estimator are based on 10 random re-orderings of the data. The results without  random re-ordering are very similar (see Figure S1 in the Supplementary Materials).

The random ordering of the samples has little affect on the  results. For $s=s_x=s_y=1$, we calculated the $95\%$ confidence intervals based on 100 random re-ordering of the original data. The endpoints of the CIs are displayed in Figure~\ref{fig:GBM100}, where each point in the scatterplot represents one CI. All the CIs have very similar widths, and are far away from zero, which is consistent with the finding of very small p-values.

Table~\ref{tab:GBM}  lists the most  correlated variables  under sparsity level  $s=3$. Interestingly, the first two microRNA measurements (\emph{hsa.miR.219} and \emph{hsa.miR.222}) also appear in a reported dependency network of important microRNAs obtained by precision matrix estimation \cite[Figure 1]{wang2015joint}. The top 25 microRNA and top 25 gene expressions in our analysis are provided in the Supplementary Materials.

\begin{table}
	\centering
	\begin{tabular}{r|rrr}
		\hline
		& hsa.miR.219 & hsa.miR.222 & hsa.miR.138 \\ 
		\hline
		SPIRE2 & 0.76 & $-$0.27 & 0.50 \\ 
		FGF9 & 0.76 & $-$0.06 & 0.33 \\ 
		ZNF553 & $-$0.24 & 0.61 & $-$0.17 \\ 
		\hline
	\end{tabular}
	\caption{GBM data analysis. Marginal correlations between the selected genes (each row) and microRNAs (each column) at sparsity level  $s=s_x=s_y=3$. The stabilized one-step estimator  $\widehat\tau_{\max} =0.931$ with standard error $0.085$.  }\label{tab:GBM}
\end{table}

\section{Discussion \label{sec:discussion} }
In this article, we  develop a new method for sparse CCA in terms of a stabilized one-step estimator for the maximal root-Pillai trace  {$\tau_{\max}$} at prespecified sparsity levels. We establish the asymptotic normality of this estimator and the validity of a confidence interval for $\tau_{\max}$ when the number of variables diverge with the sample size. Based on a greedy search algorithm, we are able to obtain a computationally tractable approximate solution that is feasible even when the pre-specified sparsity levels are moderately large. Although addressing a non-regular estimation problem, the proposed stabilized one-step estimator for the targeted maximal root-Pillai trace is asymptotically efficient when it has a unique maximizing set of indices.  Further, the asymptotic theory we develop also applies to the result of the greedy search algorithm (which targets a parameter  that is in general close to, but not identical to, the maximal root-Pillai trace).
Our simulation studies show the method performs well provided the true sparsity levels are in the range $1$--$10$, and it outperforms Bonferroni-corrected and higher criticism MANOVA F-tests.  
Difficulties occur for the proposed method with weak dense signals (many small canonical correlation coefficients) due to instability in the sample coherence matrix as the prespecified sparsity levels become relatively large ($\ge 20$ say).  


A direction for future research is the extension from linear to non-linear relationships in the setting of  model-free sufficient dimension reduction \citep{li2018sufficient}. Such an  extension is  related to the testing of predictor contributions \citep{cook2004testing} and the recent post-dimension reduction inference framework of \cite{kim2020post}. For example, the targeted covariance  matrix in sliced inverse regression \citep{li1991sliced}, namely $\Cov\{\E(\mbX\mid \mbY)\}$, could be used in a similar role as $\bolLambda_{\mbX\mbY}$ in our setting. Methods of post-selection inference have yet to be developed in these settings as far as we know.

\subsection*{Acknowledgments}
The authors thank Dr.~Aaron Molstad from the University of Florida for sharing the pre-processed Glioblastoma Multiforme data set.  IWM was supported by NIH under award 1R01 AG062401. XZ was supported by NSF under award CCF-1908969.

\appendix
\setcounter{equation}{0}
\global\long\def\theequation{A.\arabic{equation}}%

\section*{{Appendix}}
\setcounter{section}{1}

\subsection{Derivation of the canonical gradient}

In the following derivation,  we  write
 $$\Phi(P)=\Phi^d(P)=\Vert\bolLambda_{\mbX_{\calK}\mbY_{\calJ}}\Vert_{F}^{2}=\Vert\bolLambda_{\mbX\mbY}\Vert_{F}^{2}=\tr(\bolLambda_{\mbX\mbY}\bolLambda_{\mbX\mbY}^{T})=\tr(\bolSigma_{\mbX}^{-1}\bolSigma_{\mbX\mbY}\bolSigma_{\mbY}^{-1}\bolSigma_{\mbX\mbY}^{T}).$$
Using standard results from matrix calculus,
\begin{eqnarray*}
	d\Phi(P) & = & d\tr(\bolSigma_{\mbX}^{-1}\bolSigma_{\mbX\mbY}\bolSigma_{\mbY}^{-1}\bolSigma_{\mbX\mbY}^{T})\\
	& = & \dfrac{\partial\tr(\bolSigma_{\mbX}^{-1}\bolSigma_{\mbX\mbY}\bolSigma_{\mbY}^{-1}\bolSigma_{\mbX\mbY}^{T})}{\partial\bolSigma_{\mbX}}\cdot d\bolSigma_{\mbX}+\dfrac{\partial\tr(\bolSigma_{\mbX}^{-1}\bolSigma_{\mbX\mbY}\bolSigma_{\mbY}^{-1}\bolSigma_{\mbX\mbY}^{T})}{\partial\bolSigma_{\mbY}}\cdot d\bolSigma_{\mbY}\\
	& &+  \dfrac{\partial\tr(\bolSigma_{\mbX}^{-1}\bolSigma_{\mbX\mbY}\bolSigma_{\mbY}^{-1}\bolSigma_{\mbX\mbY}^{T})}{\partial\bolSigma_{\mbX\mbY}}\cdot d\bolSigma_{\mbX\mbY}\\
	& =& -\bolSigma_{\mbX}^{-1}\bolSigma_{\mbX\mbY}\bolSigma_{\mbY}^{-1}\bolSigma_{\mbX\mbY}^{T}\bolSigma_{\mbX}^{-1}\cdot d\bolSigma_{\mbX}-\bolSigma_{\mbY}^{-1}\bolSigma_{\mbX\mbY}^{T}\bolSigma_{\mbX}^{-1}\bolSigma_{\mbX\mbY}\bolSigma_{\mbY}^{-1}\cdot d\bolSigma_{\mbY}\\
	&& +  2\bolSigma_{\mbX}^{-1}\bolSigma_{\mbX\mbY}\bolSigma_{\mbY}^{-1}\cdot d\bolSigma_{\mbX\mbY},
\end{eqnarray*}
where the dot indicates inner product of two matrices of same dimension:
$\mbA\cdot d\mbB=\langle\mbA,d\mbB\rangle=\vecc^{T}(\mbA)\vecc(d\mbB)=\tr(\mbA^{T}d\mbB)$.
Hence  
\begin{eqnarray*}
	\dfrac{d\Phi(P_{\epsilon})}{d\epsilon} & = & -\tr\left\{ \bolSigma_{\mbX}^{-1}\bolSigma_{\mbX\mbY}\bolSigma_{\mbY}^{-1}\bolSigma_{\mbX\mbY}^{T}\bolSigma_{\mbX}^{-1}\dfrac{d\bolSigma_{\mbX}(P_{\epsilon})}{d\epsilon}\right\} \\
	& & - \tr\left\{ \bolSigma_{\mbY}^{-1}\bolSigma_{\mbX\mbY}^{T}\bolSigma_{\mbX}^{-1}\bolSigma_{\mbX\mbY}\bolSigma_{\mbY}^{-1}\dfrac{d\bolSigma_{\mbY}(P_{\epsilon})}{d\epsilon}\right\} 
	 + 2\tr\left\{ \bolSigma_{\mbX}^{-1}\bolSigma_{\mbX\mbY}\bolSigma_{\mbY}^{-1}\dfrac{d\bolSigma_{\mbX\mbY}(P_{\epsilon})}{d\epsilon}\right\},
\end{eqnarray*}
recalling that $P_{\epsilon}=(1-\epsilon)P+\epsilon\delta_{\mbo}$,
where $\delta_{\mbo}$ is the Dirac measure at $\mbo=(\mbx^{T},\mby^{T})^{T}$.
Expressing $\bolSigma_{\mbX}(P_{\epsilon})$ as
\[
(1-\epsilon)P(\mbX\mbX^{T})+\epsilon\mbx\mbx^{T}-(1-\epsilon)^{2}P(\mbX)P(\mbX^{T})-\epsilon(1-\epsilon)\{P(\mbX)\mbx^{T}+\mbx P(\mbX^{T})\}-\epsilon^{2}\mbx\mbx^{T},
\]
by direct calculation we then have
$$	\left.\dfrac{d\bolSigma_{\mbX}(P_{\epsilon})}{d\epsilon}\right|_{\epsilon=0}  =  \lim_{\epsilon\downarrow0}\dfrac{\bolSigma_{\mbX}(P_{\epsilon})-\bolSigma_{\mbX}(P)}{\epsilon}
	 =  \{\mbx-P(\mbX)\}\{\mbx^{T}-P(\mbX^{T})\}.
$$
Similarly, 
$$
	\left.\dfrac{d\bolSigma_{\mbX\mbY}(P_{\epsilon})}{d\epsilon}\right|_{\epsilon=0}  =  \{\mbx-P(\mbX)\}\{\mby^{T}-P(\mbY^{T})\},\ \ \ 
	\left.\dfrac{d\bolSigma_{\mbY}(P_{\epsilon})}{d\epsilon}\right|_{\epsilon=0}  =  \{\mby-P(\mbY)\}\{\mby^{T}-P(\mbY^{T})\}.
$$
Plugging-in these expressions, and including the relevant sets of indices $\calK$ and $\calJ$, we
obtain canonical gradient of $\Phi^{d}(P)$ as stated in \eqref{CanonicalGradient}:
\begin{eqnarray}
\left.\dfrac{d\Phi^{d}(P_{\epsilon})}{d\epsilon}\right|_{\epsilon=0} & = & -\{\mbx_{\calK}-\E_{P}(\mbX_{\calK})\}^{T}\bolSigma_{\mbX_{\calK}}^{-1}\bolSigma_{\mbX_{\calK}\mbY_{\calJ}}\bolSigma_{\mbY_{\calJ}}^{-1}\bolSigma_{\mbY_{\calJ}\mbX_{\calK}}\bolSigma_{\mbX_{\calK}}^{-1}\{\mbx_{\calK}-\E_{P}(\mbX_{\calK})\}\nonumber \\
&  & -\{\mby_{\calJ}-\E_{P}(\mbY_{\calJ})\}^{T}\bolSigma_{\mbY_{\calJ}}^{-1}\bolSigma_{\mbY_{\calJ}\mbX_{\calK}}\bolSigma_{\mbX_{\calK}}^{-1}\bolSigma_{\mbX_{\calK}\mbY_{\calJ}}\bolSigma_{\mbY_{\calJ}}^{-1}\{\mby_{\calJ}-\E_{P}(\mbY_{\calJ})\}\nonumber \\
&  & +2\{\mby_{\calJ}-\E_{P}(\mbY_{\calJ})\}^{T}\bolSigma_{\mbY_{\calJ}}^{-1}\bolSigma_{\mbY_{\calJ}\mbX_{\calK}}\bolSigma_{\mbX_{\calK}}^{-1}\{\mbx_{\calK}-\E_{P}(\mbX_{\calK})\}.\nonumber
\end{eqnarray}
Then, as discussed in  Section \ref{target},   the canonical
gradient of $\Psi^{d}(P)=\Vert\bolLambda_{\mbX_{\calK}\mbY_{\calJ}}\Vert_{F}$ 
is obtained via the relationship $\{\Psi^{d}(P)\}^{2}=\Phi^{d}(P)$, resulting in 
$$
D^{d}(P)(\mbo)  =  \left.\dfrac{d\Psi^{d}(P_{\epsilon})}{d\epsilon}\right|_{\epsilon=0}=\frac{1}{2\Psi^{d}(P)}\left.\dfrac{d\Phi^{d}(P_{\epsilon})}{d\epsilon}\right|_{\epsilon=0}$$
when $\Psi^{d}(P)\neq 0$.

To prove $\E_{P}\{D^{d}(P)(\mbO)\}=0$   it  suffices to show that the canonical gradient of $\Phi^{d}(P)$ has zero mean.
We apply the following property of trace and expectation operators. For any random vector $\mbX\in\mbbR^p$ with entries having finite second moments  and any non-stochastic matrix $\mbM\in\mbbR^{p\times p}$,  
\begin{equation}\label{exptrace}
\E (\mbX^T\mbM\mbX)=\E\{\tr(\mbX^T\mbM\mbX)\}=\E\{\tr(\mbM\mbX\mbX^T)\}=\tr\{\mbM\E(\mbX\mbX^T)\}.
\end{equation}
Therefore, direct calculation shows that
\begin{eqnarray}
\E\left\{ \left.\dfrac{d\Phi^{d}(P_{\epsilon})}{d\epsilon}\right|_{\epsilon=0}\right\} \! \! \! \! & = &\!  \! \! \! -\tr\left(\bolSigma_{\mbX_{\calK}}^{-1}\bolSigma_{\mbX_{\calK}\mbY_{\calJ}}\bolSigma_{\mbY_{\calJ}}^{-1}\bolSigma_{\mbY_{\calJ}\mbX_{\calK}}\bolSigma_{\mbX_{\calK}}^{-1}\cdot\E_{P}\left[\{\mbX_{\calK}-\E_{P}(\mbX_{\calK})\}\{\mbX_{\calK}-\E_{P}(\mbX_{\calK})\}^{T}\right] \right)\nonumber \\
&  &\! \! \! \! -\tr\left(\bolSigma_{\mbY_{\calJ}}^{-1}\bolSigma_{\mbY_{\calJ}\mbX_{\calK}}\bolSigma_{\mbX_{\calK}}^{-1}\bolSigma_{\mbX_{\calK}\mbY_{\calJ}}\bolSigma_{\mbY_{\calJ}}^{-1}\cdot\E_{P}\left[\{\mbY_{\calJ}-\E_{P}(\mbY_{\calJ})\}\{\mbY_{\calJ}-\E_{P}(\mbY_{\calJ})\}^{T}\right]\right)\nonumber \\
&  & \! \! \! \!+2\tr\left(\bolSigma_{\mbY_{\calJ}}^{-1}\bolSigma_{\mbY_{\calJ}\mbX_{\calK}}\bolSigma_{\mbX_{\calK}}^{-1}\cdot\E_{P}\left[\{\mbX_{\calK}-\E_{P}(\mbX_{\calK})\} \{\mbY_{\calJ}-\E_{P}(\mbY_{\calJ})\}^{T}\right] \right)\nonumber \\
&=&
-\Phi^d(P)-\Phi^d(P)+2\Phi^d(P)=0.\nonumber
\end{eqnarray}
When $\Psi^d(P)=0$, using instead the expression \eqref{newDefCG} for the canonical gradient and again applying \eqref{exptrace} we  obtain
\begin{eqnarray}
\E_P\{D^{d}(P)(\mbO) \} 
& = & \E_P\left[ \{\mbY_{\calJ}-\E_{P}(\mbY_{\calJ})\}^{T}\bolSigma_{\mbY_{\calJ}}^{-1/2}{\bf L}\bolSigma_{\mbX_{\calK}}^{-1/2}\{\mbX_{\calK}-\E_{P}(\mbX_{\calK})\}\right]\nonumber \\
& = & \tr\left( {\bf L}\bolSigma_{\mbX_{\calK}}^{-1/2} \E_P \left[ \{\mbX_{\calK}-\E_{P}(\mbX_{\calK})\}\{\mbY_{\calJ}-\E_{P}(\mbY_{\calJ})\}^{T}\right] \bolSigma_{\mbY_{\calJ}}^{-1/2}  \right) \nonumber \\
& = & \tr ({\bf L} \bolLambda_{\mbX_{\calK}\mbY_{\calJ}})=0  \nonumber\end{eqnarray}
since $\bolLambda_{\mbX_{\calK}\mbY_{\calJ}}=0$ when $\Psi^d(P)=0$.
\subsection{Proof of Lemma~\ref{lem: stepwise}}

Let $\mbbX\in\mbbR^{n\times s_{x}}$, $\mbbY\in\mbbR^{n\times s_{y}}$
and $\mbbZ\in\mbbR^{n\times1}$ be the centered data matrices of $\mbX_{\calK}$,
$\mbY_{\calJ}$ and $X_{k}$. That is, the $i$-th row of $\mbbX$
is $(\mbX_{\calK,i}-\overline{\mbX}_{\calK})^{T}$. Note that $\mbS_{\mbX_{\calK}}=\mbbX\mbbX^{T}/n$
is the sample covariance matrix of $\mbX_{\calK}$, which is assumed
to be positive definite. Then we can write $\mbC_{\mbY_{\calJ}\mbX_{\calK}}=n^{-1}\mbS_{\mbY_{\calJ}}^{-1/2}\mbbY^{T}\mbbX\mbS_{\mbX_{\calK}}^{-1/2}$
and $\mbC_{\mbY_{\calJ}\mbX_{\calK}}\mbC_{\mbY_{\calJ}\mbX_{\calK}}^{T}=n^{-1}\mbS_{\mbY_{\calJ}}^{-1/2}\mbbY^{T}\mbbX(\mbbX^{T}\mbbX)^{-1}\mbbX^{T}\mbbY\mbS_{\mbY_{\calJ}}^{-1/2}$,
where  $\mbbX(\mbbX^{T}\mbbX)^{-1}\mbbX^{T}\equiv \mbP_{\mbbX}\in\mbbR^{n\times n}$
is the projection matrix onto the $s_{x}$-dimensional subspace spanned
by the columns of $\mbbX$. Then, 
\begin{equation}
	\Vert\mbC_{\mbY_{\calJ}\mbX_{\calK}}\Vert_{F}^{2}=\tr(\mbC_{\mbY_{\calJ}\mbX_{\calK}}\mbC_{\mbY_{\calJ}\mbX_{\calK}}^{T})=n^{-1}\tr(\mbS_{\mbY_{\calJ}}^{-1/2}\mbbY^{T}\mbP_{\mbbX}\mbbY\mbS_{\mbY_{\calJ}}^{-1/2}),\label{supp_pf_lem1_1}
\end{equation}
\begin{equation}
	\Vert\mbC_{\mbY_{\calJ}\mbX_{\calK\cup k}}\Vert_{F}^{2}-\Vert\mbC_{\mbY_{\calJ}\mbX_{\calK}}\Vert_{F}^{2}=n^{-1}\tr\{\mbS_{\mbY_{\calJ}}^{-1/2}\mbbY^{T}(\mbP_{(\mbbX,\mbbZ)}-\mbP_{\mbbX})\mbbY\mbS_{\mbY_{\calJ}}^{-1/2}\},\label{supp_pf_lem1_2}
\end{equation}
where $\text{\ensuremath{\mbP_{(\mbbX,\mbbZ)}}\ensuremath{\ensuremath{\in\mbbR^{n\times n}}}}$
is the projection matrix onto the $(s_{x}+1)$-dimensional subspace of $\mbbR^{n}$
spanned by the columns of $(\mbbX,\mbbZ)$. That is, 
\begin{eqnarray*}
	\mbP_{(\mbbX,\mbbZ)} & = & \left(\begin{array}{cc}
		\mbbX & \mbbZ\end{array}\right)\left(\begin{array}{cc}
		\mbbX^{T}\mbbX & \mbbX^{T}\mbbZ\\
		\mbbZ^{T}\mbbX & \mbbZ^{T}\mbbZ
	\end{array}\right)^{-1}\left(\begin{array}{c}
		\mbbX^{T}\\
		\mbbZ^{T}
	\end{array}\right)\\
	& = & n^{-1}\left(\begin{array}{cc}
		\mbbX & \mbbZ\end{array}\right)\left(\begin{array}{cc}
		\mbS_{\mbX_{\calK}}^{-1}+\mbS_{\mbX_{\calK}}^{-1}\mbS_{\mbX_{\calK}X_{k}}\mbS_{R_{k}}^{-1}\mbS_{X_{k}\mbX_{\calK}}\mbS_{\mbX_{\calK}}^{-1} & -\mbS_{\mbX_{\calK}}^{-1}\mbS_{\mbX_{\calK}X_{k}}\mbS_{R_{k}}^{-1}\\
		-\mbS_{R_{k}}^{-1}\mbS_{X_{k}\mbX_{\calK}}\mbS_{\mbX_{\calK}}^{-1} & \mbS_{R_{k}}^{-1}
	\end{array}\right)\left(\begin{array}{c}
		\mbbX^{T}\\
		\mbbZ^{T}
	\end{array}\right)\\
	& = & n^{-1}\mbbX\mbS_{\mbX_{\calK}}^{-1}\mbbX^{T}+n^{-1}(\mbbZ-\mbbX\mbS_{\mbX_{\calK}}^{-1}\mbS_{\mbX_{\calK}X_{k}})\mbS_{R_{k}}^{-1}(\mbbZ^{T}-\mbS_{X_{k}\mbX_{\calK}}\mbS_{\mbX_{\calK}}^{-1}\mbbX^{T}),\\
	& = & \mbP_{\mbbX}+\mbP_{\mbbZ\mid\mbbX},
\end{eqnarray*}
where $\mbS_{R_{k}}=n^{-1}\{\mbbZ^{T}\mbbZ-\mbbZ^{T}\mbbX(\mbbX^{T}\mbbX)^{-1}\mbbX^{T}\mbbZ\}$
is the sample covariance of the fitted residual $(\mbbZ-\mbbX\mbS_{\mbX_{\calK}}^{-1}\mbS_{\mbX_{\calK}X_{k}})$.
Specifically, $(\mbbZ-\mbbX\mbS_{\mbX_{\calK}}^{-1}\mbS_{\mbX_{\calK}X_{k}})^{T}(\mbbZ-\mbbX\mbS_{\mbX_{\calK}}^{-1}\mbS_{\mbX_{\calK}X_{k}})=n\mbS_{R_{k}}$.
Therefore, $\mbP_{\mbbZ\mid\mbbX}\in\mbbR^{n\times n}$ is indeed
the projection matrix onto the one-dimensional subspace spanned by
$(\mbbZ-\mbbX\mbS_{\mbX_{\calK}}^{-1}\mbS_{\mbX_{\calK}X_{k}})$.
The conclusion follows from \eqref{supp_pf_lem1_1}, \eqref{supp_pf_lem1_2}
and $\mbP_{(\mbbX,\mbbZ)}=\mbP_{\mbbX}+\mbP_{\mbbZ\mid\mbbX}$.

\subsection{Recursive computation of the one-step estimator \label{sec:recur}}
In the implementation, given a new observation $\mbO_{j+1}$, we need to update the weight $w_j=\overline\sigma_n/\widehat\sigma_j$ and the variance of the canonical gradient $\widehat\sigma_j^2$ in order to update the current estimate $\psi_j$ of the target parameter $\tau_{\max}$. Since we only need to know these quantities when all observations are included, an  efficient approach is to exploit the following recursive properties of $\psi_j/\overline{\sigma}_j$ and $\overline\sigma_j^{-1}$:\begin{eqnarray*}
	\dfrac{\psi_{j+1}}{\overline{\sigma}_{j+1}} & = & \dfrac{1}{j+1}\left\{ j\cdot\dfrac{\psi_{j}}{\overline{\sigma}_{j}}+\dfrac{\Psi^{d_{nj}}(P_{j})+D^{d_{j}}(P_{j})(\mbO_{j+1})}{\widehat{\sigma}_{nj}}\right\} ,\ \ \ 
	\overline{\sigma}_{j+1}^{-1}  =  \dfrac{1}{j+1}\left(j\cdot\overline{\sigma}_{j}^{-1}+\widehat{\sigma}_{nj}^{-1}\right).
	\end{eqnarray*}
We then obtain the final estimate when $j=n$:
$$
\widehat\tau_{\max} \equiv \psi_{n}  =   \dfrac{\overline{\sigma}_{n}}{n-\ell_{n}}\sum_{j=\ell_{n}}^{n-1}\left\{ \dfrac{\Psi^{d_{nj}}(P_{j})+D^{d_{j}}(P_{j})(\mbO_{j+1})}{\widehat{\sigma}_{nj}}\right\},\ \ \ \overline{\sigma}_{n}^{-1}  =  \dfrac{1}{n-\ell_{n}}\sum_{j=\ell_{n}}^{n-1}\widehat{\sigma}_{nj}^{-1}.
$$

\baselineskip=14pt  
\bibliographystyle{apalike}
\bibliography{TestingCCA}

\end{document}